%% file: warpn.tex
\input macros.tex

\input jour.tex

\input psfig 

\overfullrule=0pt
\magnification 1200
\baselineskip=15 true bp

\centerline{\bf On War: The Dynamics of Vicious Civilizations}
\bigskip\bigskip\bigskip
\centerline{I.~Ispolatov$^1$, P.~L.~Krapivsky{$^2$}, and S.~Redner$^1$}
\bigskip 
\centerline{$^1$\sl Center for Polymer Studies and Department of Physics}
\centerline{\sl Boston University, Boston, MA 02215}
\medskip
\centerline{$^2$\sl Courant Institute of Mathematical Sciences}
\centerline{\sl New York University, New York, NY 10012-1185}

\vskip 0.3in

\centerline{\bf Abstract}

{\narrower\smallskip\noindent The dynamics of ``vicious'',
continuously growing civilizations (domains), which engage in ``war''
whenever two domains meet, is investigated.  In the war event, the
smaller domain is annihilated, while the larger domain is reduced in
size by a fraction $\e$ of the casualties of the loser.  Here $\e$
quantifies the fairness of the war, with $\e=1$ corresponding to a fair
war with equal casualties on both side, and $\e=0$ corresponding to a
completely unfair war where the winner suffers no casualties.  In the
heterogeneous version of the model, evolution begins from a specified
initial distribution of domains, while in the homogeneous system, there
is a continuous and spatially uniform input of point domains, in
addition to the growth and warfare.  For the heterogeneous case, the
rate equations are derived and solved, and comparisons with numerical
simulations are made.  An exact solution is also derived for the case of
equal size domains in one dimension.  The heterogeneous system is found
to coarsen, with the typical cluster size growing linearly in time $t$
and the number density of domains decreases as $1/t$. For the
homogeneous system, two different long-time behaviors arise as a
function of $\e$.  When $1/2<\e\leq 1$ (relatively fair wars), a steady
state arises which is characterized by egalitarian competition between
domains of comparable size.  In the limiting case of $\e=1$, rate
equations which simultaneously account for the distribution of domains
and that of the intervening gaps are derived and solved.  The steady
state is characterized by domains whose age is typically much larger
than their size.  When $0\leq\e<1/2$ (unfair wars), a few
``superpowers'' ultimately dominate.  Simulations indicate that this
coarsening process is characterized by power-law temporal behavior, with
non-universal $\e$-dependent exponents.  Some of these features are
captured by a deterministic self-similar model, for which the
characteristic exponents can be computed easily.  The transition point
$\e=\e_c=1/2$ is characterized by slower than power-law coarsening.

\bigskip
\bigskip\noindent
P. A. C. S. Numbers:  01.75+m, 02.50.-r, 89.90+n

}
\baselineskip=17.5 true bp

\vfill\eject

\rchap{INTRODUCTION}

\asubsec{Background} 

Coarsening phenomena underlie a wide variety of physical processes, such
as phase ordering and spinodal decomposition\ref{1-3}, growth of breath
figures\ref{4,5}, spin dynamics\ref{6}, and foams\ref{7,8}.  The latter
system is especially interesting, as the basic phenomena are readily
observable in many everyday situations.  The microscopic rules which
govern the dynamics of individual bubbles are simply given in terms of
the geometry and yet lead to varied and intriguing macroscopic behavior
in the long-time limit.  In addition to continuous growth and shrinking
of individual bubbles, there are discontinuous bubble ``popping'' events
which lead to rearrangement on a larger scale than individual bubble
growth.  Another attractive feature of such geometric coarsening
processes is that they naturally suggest idealizations which may be
exactly soluble.  One such example, which can be viewed as a limiting
case for breath figure growth, is the coarsening of an array of
contiguous domains in one dimension\ref{9-17}.

The mechanisms that govern domain evolution in these types of coarsening
processes appear to have natural counterparts in social phenomena.  For
example, the competition between cultures has led to a rich historical
record\ref{18} in which certain civilizations are dominant for long time
periods only to suddenly disappear.  Conversely, other civilizations
persist for very long times even though they are relatively small.
Motivated by these basic historical facts and by a qualitative
appreciation for coarsening phenomena, we introduce and investigate a
simple, yet relatively general model for war between vicious
civilizations.  In our model, domains grow continuously and an encounter
between two civilizations leads to a war where the smaller combatant is
annihilated, and the larger civilization suffers a specified number of
casualties.  This model exhibits a rich variety of dynamical behaviors,
including the features of extinction and persistence of civilizations.

\asubsec{The War Model} In our model, space is populated by a collection
of civilizations, each of which is represented by a spatial domain of a
particular size (or population).  Each domain grows either continuously
or in discrete steps at a constant rate, in which the boundaries move at
velocity $V$.  Whenever two civilizations of size $i$ and size $j$ meet
(with $j>i$ without loss of generality), they engage in a war where the
smaller domain is annihilated, while the larger domain suffers $\e\, i$
casualties, so that its size changes to $j-\e\,i$ (Fig.~1).  We define
the casualties to occur at the battlefront so that the frontier of the
winner retreats by a distance $\e\,i$ in one dimension.  Here $\e$
measures the ``fairness'' of the warfare event; $\e=1$ corresponds to a
fair war, in which the winner and loser suffer the same number of
casualties, and $\e=0$ corresponds to a completely unfair war in which
the winner emerges unscathed.  Subsequently, the survivor civilization
continues its growth until the next war.  We are interested in
determining the long-time dynamical behavior and the spatial
distribution of civilizations under these conditions.  Although our
model is naive and drastic (since the loser is annihilated), the
essential features of growth and sudden diminishment by war are
incorporated.  Thus suitable generalizations of our model may be
appropriate for describing quantitative aspects of social history.

There are two general conditions under which our model leads to
interesting asymptotic dynamics.  In the {\it heterogeneous} warfare
process, the system begins with a distribution of nascent domains which
subsequently undergo growth and warfare.  In the {\it homogeneous}
process, there is temporally constant and spatially uniform input of
point domains which subsequently grow and experience wars whenever two
domains meet.  From a social perspective, this input could be viewed as
arising from the remnants of destroyed civilizations.

\asubsec{Overview of Dynamical Properties} 

For the heterogeneous warfare process, we have primarily concentrated on
the case $\e=1$, although qualitatively similar results are anticipated
for any value of $\e$.  This system exhibits coarsening, in which the
domain size distribution approaches a scaling form in $x/t$, where $x$
is the domain size and $t$ is the time.  Consequently the average domain
size, $\av{x(t)}$, grows linearly in time.  These qualitative results
hold both in the mean-field limit and in one dimension.  From the rate
equations, the scaling function for the domain size distribution is
one-half of a sinusoid.  In contrast, simulations in one dimension
reveal a scaling function for the size distribution which is triangular
in shape.

The homogeneous warfare process exhibits a richer phenomenology which is
controlled by the fairness parameter $\e$.  For $1/2<\e\leq 1$, a
steady-state arises, which is characterized by egalitarian competition
between domains of similar size, while for $\e<1/2$, coarsening occurs
in which a few superpowers ultimately dominate.  The existence of a
threshold between these two regimes can be qualitatively justified by
considering the outcome of a war between two neighboring domains with
different birth times (or equivalently, sizes).  For $\e_c=1/2$, when
two neighboring domains meet and engage in war, the frontier of the
survivor after the war will retreat to exactly the same position that
existed at the birth-time of the smaller civilization (Fig.~2).  Thus
for $\e\gl 1/2$, the frontier of the victor after the warfare event
should either advance or retreat, respectively, compared to its position
at the birth-time of the weaker combatant.  This difference suggests the
existence of the aforementioned steady state and coarsening regimes
which are separated by $\e_c= 1/2$.

An interesting feature of the steady state regime, $\e>\e_c$, is that a
non-trivial joint distribution of civilization age and size arises.  The
connection between these two attributes is subtle, since a long-lived
civilization, whose age is much greater than its size (in appropriately
scaled units where $V=1/2$), can arise by a sequence of fortuitous
events.  To characterize this age-size distribution, we have formulated
a version of the rate equations which simultaneously accounts for the
distribution of domains, as well as the distribution of gaps between
domains.  This approach appears to provide some exact results for the
age-size distribution in one dimension, at least in the analytically
tractable case of $\e=1$.  For the distribution of domain sizes
(integrated over all domain ages), $n(x)$, we find $n(x)=e^{-x/x_0}$
with $x_0=1/\sqrt{\mu}$, in excellent agreement with simulations.  Here
$\mu$ is the rate per unit length at which new civilizations are
introduced.  More interestingly, the age distribution (integrated over
all domain sizes), $n(\tau)$, has an exponential tail, $n(\tau)\propto
e^{-\t/\t_0}$ as $\t\to\infty$, but with an anomalously large
characteristic age $\t_0$ which is much larger than the typical size.
Simulations in one dimension give an even larger value for the ratio
$\t_0/x_0$, a result which can be attributed to the existence of
significant anti-correlations in the ages of neighboring domains.  For
the case of general $\e$ above the threshold, $\e_c<\e\leq 1$, our
simulations indicate that basic quantities, such as the average domain
size, domain age, and coverage, all eventually reach steady-state
values.  Correspondingly, both the domain size and age distributions
decay exponentially in this regime.

In the complementary regime of $\e<\e_c$, warfare events sufficiently
favor the victor that superpowers ultimately emerge.  The existence of
these large domains strongly modifies the effect of the continuous input
of small domains, so that a steady state is not achieved.  Our
simulations indicate a continuous coarsening of domains in which basic
time-dependent quantities, such as the mean domain size and the number
density of domain, exhibit non-universal $\e$-dependent power law
behavior in time.  In particular, for the extreme limit of $\e=0$, where
the victor in a war suffers no casualties, the number of domains and the
density of empty space decay as $t^{-\g}$ in one dimension, with
$\g\approx 1/3$.  A deterministic idealization of the war model is
introduced which provides a useful description of this coarsening
phenomenon.  As $\e$ approaches $\e_c$ from below, coarsening becomes
slower, as evidenced by the $\e$ dependence of the exponents.  For
$\e\approx \e_c$, marginal behavior occurs in which the average domain
size grows extremely slowly, while the average age continues to grow as
a power law in time.

In Sec.\ II, we study the heterogeneous war model with fair wars,
$\e=1$.  Results from a mean-field theory and numerical simulations are
presented.  We also treat the special case where domains all have the
same initial size, so that both domains disappear in a warfare event.
This limiting situation turns out to be exactly soluble in one dimension
for both the asymptotic value of the coverage, as well as the domain
size distribution.  We next turn to the homogeneous process in Sec.\
III.  From the rate equations, the equation of the steady-state solution
for the joint age-size distribution is obtained, again for the
particular case of fair war, $\e=1$.  The extension of the rate
equations for $\e<1$ is also discussed and an argument for the existence
of a transition in the kinetic behavior at $\e=\e_c$ is presented.  We
next present simulation results in one dimension, focusing on general
properties as a function of the fairness parameter $\e$.  Basic features
of the $\e>\e_c$ steady-state regime and the $\e <\e_c$ coarsening
regime are outlined.  A deterministic self-similar model is then
introduced to help understand the coarsening process for $\e=0$.  In
Sec.\ IV, we give a brief summary and discuss several extensions of our
model.  Various calculational details are given in the Appendices.

\rchap{HETEROGENEOUS WAR}\newcount\nlet\nlet=0

\asubsec{Mean-Field Theory} 

Consider the mean-field limit, in which pairs of domains are randomly
picked to undergo warfare.  With this interaction rule, the evolution of
the domain size distribution is described by the rate equations
\ref{19},
$$
\dot c_k(t)=\sum_{i=1}^\infty c_i(t)c_{i+k}(t)- c_k(t)\sum_{i=1}^\infty c_i(t)
+\l(c_{k-1}(t)-c_k(t)).
\eqnoi 
$$ 
Here $c_k(t)$ is the concentration of domains of size $k$ at time $t$.
The first term accounts for the gain of $k$-domains due to a war between
domains of size $i$ and $i+k$.  Similarly, the second term accounts for
the loss of $k$-domains because of war between a domain of size $k$ and
any other domain.  The gain and loss of $k$-domains due to constant
growth at rate $\l$ are described by the last two terms.  We have
implicitly assumed that the warfare rate is independent of the combatant
sizes.  \last\ also assumes a minimal size of unity, so that all sizes
are integers.  The extension to the continuum case is straightforward;
and this description will be employed for the homogeneous war model.

For the system described by \last, the average civilization size grows
linearly with time and the size distribution is one-half of a sinusoid.
To derive these results, we first identify the appropriate scaling
variable.  For this purpose, let us temporarily neglect the effect of
war.  The size distribution for such peacefully flourishing
civilizations is found from the rate equation $\dot
c_k=\l(c_{k-1}-c_k)$.  This is just the Poisson process, with solution,
for the ``Adam'' initial condition of $c_k(t=0)=\delta_{k,1}$,
$$
c_k(t)={(\l t)^{k-1}\over (k-1)!}e^{-\l t}
\sim {1\over \sqrt{2\pi\l t}}\exp\left[-{(k-\l t)^2\over \l t}\right]. 
\eqnoi
$$
While the exact result is specific for the monodisperse initial
condition, the Gaussian approximation provides universal asymptotics for
an arbitrary compact initial size distribution.  Thus for peaceful
civilizations, the size distribution is peaked around $k=\l t$, with
dispersion $\sqrt{\l t}$.  Clearly, warfare produces sizes with $k<\l
t$, so that the size distribution for warring civilizations should be
non-vanishing for all $k\leq \l t$.  This suggests that the appropriate
scaling variable is $x=k/\l t$, with $0\leq x \leq 1$.  The leading
region where $k\approx\l t$ can be expected to have a fine structure
over an extent of the order of $\sqrt{\l t}$, as in the special case of
peaceful civilizations.  In the following, we ignore this detailed
structure.

To solve the rate equations, it proves useful to consider first the
civilization number density, $N(t)=\sum_{k\geq 1} c_k(t)$, which, from
\last, satisfies
$$
\dot N(t)=-{1\over 2}\sum_{k=1}^\infty c_k(t)^2-{1\over 2}N(t)^2.
\eqnoi
$$
Asymptotically, the first term on the right-hand side is clearly
negligible; therefore, as $t\to\infty$, $N(t)\sim 2/t$.  We thus
expect that scaling form for the civilization size distribution is
$$
c_k(t)={1\over \l t^2}~\CC(x), \qquad{\rm with}\quad x={k\over \l t}.
\eqnoi
$$
The time dependent prefactor in \last\ guarantees that $N(t)\propto
t^{-1}$.  Furthermore, the relation $N(t)\sim 2/t$ is quantitatively
satisfied if $\int_0^1 dx\,\CC(x)=2$.

Substituting the scaling ansatz into the rate equations, one finds that
in the continuous limit, the scaling function satisfies an
integro-differential equation of an anti-convolution form,
$$
(1-x)\,\CC'(x)=\int_0^{1-x}dy\,\CC(y)\,\CC(x+y),
\eqnoi
$$ 
where the prime denotes the differentiating with respect to $x$.  We are
unable to solve this equation by a systematic approach, and therefore
resort to trial-and-error.  Since $\CC'>0$, $\CC(x)$ is monotonically
increasing in $x$.  On physical grounds, we expect that $\CC(0)=0$,
while analysis of \last\ in the vicinity of $x=1$ gives
$$
\CC(x)=\CC(1)-(1-x)^2{\CC(1)\,\CC'(0)\over 4}+\ldots,
\eqnoi
$$
\ie, $\CC'(1)=0$ and $\CC''(1)<0$.  Polynomial test functions fail to
satisfy \back1. (For a polynomial, say, of degree $g$, the left-hand
side of \back1\ is a polynomial of degree $g$ while the right-hand
side has degree $2g+1$).  We find, however, that the simplest
appropriate transcendental test function, $\CC(x) =\pi~\sin\left({\pi
x\over 2}\right)$, solves \back1.  (The constants in this expression
for $\CC(x)$ are chosen to satisfy $\CC'(1)=0$ and the sum rule for
$\CC(x)$.)~ Combining \last\ with the scaling form of \back2\ gives
the basic result,

$$ 
c_k(t)\simeq\cases{{\pi\over\l t^2}\sin\left({\pi k\over 2\l
t}\right), \qquad &$k\leq \l t$, \cr\cr 0, &$k>\l t$. \cr} 
\eqnoi 
$$

{}From this solution,  the moments of the size
distribution, $M_n(t)\equiv\sum_{k\ge 1} k^n c_k(t)$ are 
given by
$$
\eqalign{
M_0&={2\over t}, \quad M_1={4\l\over \pi}, \quad
M_2={8(\pi-2)\over \pi^2}\l^2 t,~\ldots, \cr
M_n&=\left({2^{n+1}\over \pi^n}\int_0^{\pi/2}dx~x^n\sin x\right)
\l^n t^{n-1}.\cr}
\eqnoi
$$
Thus, \eg, $M_0(t)$ is the civilization number density, $M_1(t)$
is the mass density, or coverage, $M_1(t)\equiv M(t)$; \etc.
{}From \last, any reasonable measure of the typical domain size, \eg,
$[M_n(t)/M_0(t)]^{1/n}$, increases linearly with time.

\asubsec{Simulation Results in One Dimension}

A straightforward way to simulate our warfare model in one dimension is
by molecular dynamics.  Within a continuum and deterministic description
for domain growth, one identifies the minimum of the conflict times
between all pairs of nearest-neighbor domains at any given stage.  The
system evolves freely until this minimum conflict time, at which point a
war occurs, where one domain (the smaller) disappears and its (larger)
nearest neighbor shrinks.  After this event, all pairwise conflict times
are recomputed and the overall update process is repeated.  While simple
to implement, this molecular dynamics is relatively inefficient, as the
computation time is proportional to the square of the number of domains.
We therefore employed an alternative algorithm which leads to a savings
of almost two orders of magnitude in time for a system with $10^5$
initial civilizations, compared to molecular dynamics\ref{20}.  In our
approach, we first determine the conflict times for all neighboring
civilizations and sort them in ascending order.  Instead of re-computing
conflict times after each war, we continue to use the pre-sorted times
for carrying out successive warfare events until a threshold is reached.
This threshold is determined by first computing the new nearest-neighbor
conflict time that is created as a result of the current war and
comparing this new time, as well as the nearest-neighbor conflict times
which are ``lost'' by the current war, with the next conflict time on
the pre-sorted list.  If any of these putative times are less than this
next pre-sorted time, an inconsistency would arise at the next step.  It
is then necessary to re-compute and re-order all conflict times.  This
exhaustive molecular dynamics step needs to be performed relatively
rarely, leading to considerable saving in simulation time.  It is
possible, at the expense of algorithmic simplicity, to eliminate the
molecular dynamics step entirely by constantly ordering the list of
conflict times as conflict times are created and destroyed in each
warfare event.

For a polydisperse initial distribution of civilization sizes, our
simulations show that the time-dependent size distribution evolves to a
nearly universal scaling form in the asymptotic limit.  Details of the
initial condition are irrelevant as long as they are not singular in
character.  Our numerical results are based on using a Poisson initial
distribution for both the sizes of civilizations and the intervening
gaps.  However, the shape of the size distribution is influenced by
details of the warfare event, such as the location of the removed
portion of the victorious domain.  In our simulations, this removed
portion is adjacent to the battlefront (Fig.~1).  Other rules are
possible and perhaps natural, \eg, one could define a rule in which the
center-of-mass of the survivor remains fixed after the warfare event.
Because of this detail dependence of the size distribution, its
quantitative characterization is of limited value.  For the casualty
rule adopted here, the distribution has simple triangular shape
(Fig.~3).

An interesting feature from the simulations is that the sizes of nearby
domains are virtually uncorrelated.  That is, the size correlation
function, $C_s(r)\equiv\av{s_is_{i+r}}/\av{s_i}^2-1\approx 0$ for $r>1$,
where $s_i$ is the size of the \ith\ domain.  For $r=1$, the simulations
give $C_s(1)\approx -0.01$ which is at least 3 times larger than the
correlation function for any other value of $r$.  Thus two large
civilizations are less likely to coexist peacefully as nearest
neighbors; rather, a large domain is slightly more likely to be
surrounded by small neighbors and vice versa.

\asubsec{Equal Size Domains}

For heterogeneous war in one dimension, the case of initial equal-size
civilizations is unique because both combatants are eliminated in a war
and the equal-size distribution is preserved.  This case turns out to be
exactly soluble by appealing to a connection with domain coarsening
processes (see, \eg, Refs.~9-13), and generalizing the approaches in
Refs.~14-17.  In domain coarsening, which we may view as being ``dual''
to war for equal size domains (Fig.~4), the system consists of
contiguous domains of arbitrary sizes and coarsening occurs by
successive elimination of the smallest domain.  When the walls
associated with this minimal size domain disappear, the other walls
remain fixed.  This successive domain elimination corresponds exactly to
the pairwise annihilation of the two closest domains in the war model,
as illustrated in Fig.~4.  Note also that since all civilizations have
the same size, their growth rate is immaterial, and models with
size-dependent growth may be solved by the same approach as that used
for size independent growth by using the domain length $L$ as the time
parameter.

Let $n(l,L)\,dl$ be the number of neighboring civilizations of size $L$
whose centers are separated by a distance which is within $[l,l+dl]$.
Using the equivalence to coarsening, we term the interval between the
centers of neighboring civilizations a ``domain''.  The total number of
such surviving domains is
$$
\CN(L)=\int_L^\infty  n(l,L)\,dl.
\eqnoi
$$
It proves useful to normalize this quantity, $f(l,L)=n(l,L)/\CN(L)$, and
then define the (almost) scaling form, $F(x,L)=L\,f(l,L)$, with $x=l/L$.
The absence of correlations\ref{15-17} between domains in the dual
coarsening process is crucial since it implies that mean-field rate
equation for $F(x,L)$ is exact.  This rate equation
reads\ref{14}
$$ 
L\pd{}{L}F(x,L)=F(x,L)+x\pd{}{x}F(x,L)+
\theta(x-3)G(L)\int_1^{x-2}dy F(y,L)F(x-y-1,L),
\eqnoi
$$ 
where $G(L)\equiv F(x=1,L)$.  \last\ can be obtained by a
straightforward enumeration of the outcomes that arise from the
elimination of the smallest domains (see, \eg, the derivation of Eq.~(9)
in Ref.~17).  For example, the last term in the right-hand side of
\last\ describes the formation rate of an $x$-domain by elimination of
the smallest domain, of scaled length 1, which is situated between two
domains of scaled lengths $y$ and $x-y-1$.  The step function
$\theta(x-3)$ ensures that the resulting domain will be at least 3 times
larger than the minimal domain.

{}From the solution to \last\ in the long-time limit, or equivalently,
$L\to \infty$, the asymptotic coverage is 
$$
M_\infty={1\over 2 e^\g}\cong 0.28073,
\eqnai
$$
where $\g\cong 0.5772156$ is Euler-Masceroni constant.  (This asymptotic
solution, first given in Ref.~14, as well as the full time-dependent
solution are detailed in Appendix A).  Additionally, the number density
$N(L)$ of these equal-size civilizations asymptotically is
$$
N(L)\sim M_\infty L^{-1}.
\eqnoi
$$
These behaviors for the number density and the coverage are
qualitatively similar to the corresponding mean-field results.
Additionally, from the complete time-dependent solution of the rate
equations in Appendix A, one also finds the 
asymptotic expansion of the coverage,
$$
M(L)={1\over 2 e^\g}+ {A\over L}+\ldots,\eqno(11b)
$$
\ie, the first dominant correction decays as $L^{-1}$. 
Here $L$ is a measure of the physical time, and the coefficient $A$
depends on the details of the initial size distribution.

While it would be interesting to investigate the domain evolution in
heterogeneous was for general $0\leq\e<1$, the equal-size property is
lost as the process develops, and an exact treatment does not seem
possible.  However, the extreme case of completely unfair war, $\e=0$,
still enjoys the property that an equal-size distribution remains
invariant during the evolution if one defines that one of the combatants
(picked randomly) is annihilated in a war while the other remains
unchanged.  Such a model may be solvable by techniques similar to those
employed in the above case of equal size domains engaging in fair war.

\rchap{HOMOGENEOUS WARFARE}\newcount\nlet\nlet=0

We now consider the effect of a temporally and spatially homogeneous
input of size-less civilizations on the dynamics.  As discussed in the
Introduction, two fundamentally different long time behaviors can occur,
depending on the value of the fairness parameter $\e$.  For $0\leq
\e<1/2$, power-law coarsening occurs, leading to the emergence of a few
superpowers.  Conversely, for $1/2<\e\leq1$, a steady state arises, with
egalitarian competition between comparable size domains.  For the latter
situation, it is plausible that a mean-field approach might be accurate,
since the input leads to a well-mixed state.  The rate equations of the
previous section are not suitable, however, since the restriction to
nearest-neighbor interactions is not accounted for.  Our goal here is to
construct rate equations for the driven one-dimensional process which
incorporates the obvious restrictions associated with one spatial
dimension.  We are able to solve for the steady state of these governing
equations in the fair war case of $\e=1$.

\asubsec{Rate Equations for Fair Wars}

In one dimension, civilizations are represented by non-overlapping
intervals, with the civilization size equal to the interval length.  It
is now convenient to assume continuous and deterministic civilization
growth in which boundaries move with constant velocity $V$.  Thus the
random birth times and placement of new civilizations (whose initial
size may be taken as zero) are the only sources of randomness. 

Consider the fair war case of $\e=1$.  To write the rate equations, we
first introduce the distribution functions, $n(x,t)$ and $m(x,t)$, which
are, respectively, the density of civilizations of size $x$ and the
density of inter-civilization gaps of size $x$ at time $t$.  The number
density of civilizations can be written equivalently as
$$ 
N(t)=\int_0^\infty dx\, n(x,t)=
\int_0^\infty dx\, m(x,t).
\eqnoi
$$
Thus the fraction of covered space is $M(t)\equiv\int_0^\infty dx\,
x\,n(x,t)$, while the fraction of empty space is $E(t)\equiv\int_0^\infty dx\,
x\,m(x,t)$, with $M(t)+E(t)= 1$.

The rate equations for $n(x,t)$ and $m(x,t)$ are
$$
\left(\pd{}{t}+2V\pd{}{x}\right)n(x,t)=
4Vm_0(t)\left[\int_0^\infty dy\,{n(y,t)\over N(t)}\,{n(x+y,t)\over N(t)}
-{n(x,t)\over N(t)}\right]+\mu \delta(x)E(t),
\eqnai
$$
$$
\eqalign{
\left(\pd{}{t}-2V\pd{}{x}\right)m(x,t)&=
2\mu \int_x^\infty dy\, m(y,t)-\mu x m(x,t)-{2Vm_0(t)\,m(x,t)\over N(t)} \cr 
&+{2Vm_0(t)\over N^3(t)}
\int_0^x dz\, m(z,t)n\left((x-z)/2,t\right)
\int_{(x-z)/2}^\infty dy\,n(y,t).\cr}
\eqnbi
$$
In these equations, $m_0(t)\equiv m(x=0,t)$ is the density of gaps of
size zero and $\mu$ is the birth rate of new domains per unit length.
The spatial derivative term in these equations accounts for the
continuous growth of civilizations, in \lasta, and the shrinking of
gaps, in \lastb.  The right-hand sides account for the evolution as a
result of interactions (Fig.~5).  The first term on the right-hand side
of \lasta\ gives the production rate for domains of length $x$ as a
result of a war between domains of size $y$ and $x+y$.  Such an event
occurs only when the gap between these two domains vanishes -- hence the
factor $m_0(t)$.  In the mean-field approximation, the rate for this
process is proportional to the product of $m_0(t)$ and the probabilities
${n(y,t)/N(t)}$ and ${n(x+y,t)/ N(t)}$; the factor $4V$ accounts for the
two possible locations of the combatants, $(y,x+y)$ and $(x+y,y)$, times
the rate $2V$ at which the gap vanishes.  The second term on the
right-hand side of \lasta\ accounts for wars between an $x$-domain and
an arbitrary size right- or left-neighbor.  The last term gives the rate
at which size-less civilizations are created in empty space.  Only this
last term is exact \apr, because there is no factorization of
multi-particle correlation functions into single particle densities.

The terms on the right-hand side of \lastb\ are explained similarly.
The first two terms arise from the ``fragmentation'' of an empty
interval due to the input of new civilizations.  The gain term accounts
for the production of an $x$-gap due to the two ways in which a $y$-gap
can be broken into a gap size $y-x$ and $x$ by the input.  The second
term accounts for the loss of $x$-gaps due to their total rate of
breakup as a result of the input.  These two terms are, again,
presumably exact.  The last two terms describe how $m(x,t)$ evolves by
war.  The loss term arises because a war, which is adjacent to an
$x$-gap, leads to the removal of the $x$-gap if the adjoining
civilization is the loser.  There is a cancelation of a factor of $2$,
to account for the two possible locations of the warfare event, with a
factor of $1/2$, to account for the possibility that the loser may not
be adjacent to the gap.  The total rate for any war, independent of the
size of the combatants, is simply $2Vm_0(t)$.  Finally, the gain term
arises from wars between a civilization of size $(x-z)/2$ and
$y>(x-z)/2$ which is adjacent to a $z$-gap.  Since there are $x-z$
casualties in the war, the initial gap of size $z$ grows to size $x$.

\asubsec{Steady State Properties for Fair Wars}

To determine the steady state properties of these equations, it is first
helpful to consider the rate equation for the total number density of
civilizations, 
$$
\td{N(t)}{t}=-2Vm_0(t)+\mu E(t). \eqnoi
$$
This equation is exact; it can be derived directly on physical 
grounds and also follows by integrating
\backa1\ over all $x$.  In the steady state, \last\ becomes $2Vm_0=\mu E$.
Substituting this into the steady-state version of \backa1\ gives
$$ 
\td{}{x}n(x)= 2m_0\left[\int_0^\infty dy\,{n(y)\over N}\,{n(x+y)\over N} 
- -{n(x)\over N}\right]+m_0\delta(x). \eqnoi
$$

We seek an exponential solution to this equation.  The presence of the
$\delta$-function implies that $n(x=0)=m_0$.  Thus we hypothesize that
$n(x)=m_0e^{-ax}$.  It is easily verified that this satisfies \last.
Additionally, from $2Vm_0=\mu E=\mu\int_0^\infty x\,m(x)\,dx$ and
$n(x)=m_0e^{-ax}$, the parameter $a$ is determined by
$2Vm_0=\mu\left(1-{m_0\over a^2}\right)$.  To complete the solution, we
now consider the rate equation for $m(x)$.  It is again natural to
attempt the same exponential form for the gap distribution,
$m(x)=m_0e^{-ax}$.  Substituting this ansatz into the steady version of
\backb2, consistency is achieved if $\mu=2Va^2$.  Combining with the
previous relation gives $m_0=\mu/4V$.  Thus we finally arrive at the
steady state solution
$$ 
n(x)=m(x)={\mu\over 4V}\exp\left(-x\sqrt{{\mu\over 2V}}\right). \eqnoi
$$
This gives the steady state number density, $N=\sqrt{\mu/8V}$ and
coverage, $M=E=1/2$.  Results from our numerical simulations of the
homogeneous warfare model in one dimension are indistinguishable from
these results, suggesting that this rate equation approach gives the
exact the size distribution.

Consider now the civilization age distribution.  This turns out to be a
considerably more interesting but subtle characteristic of the steady
state.  Although the age distribution is asymptotically an exponentially
decaying function of age, the characteristic age is much larger than the
naive expectation of the characteristic size divided by $V=1/2$.  Thus a
typical civilization survives many wars before it is ultimately
extinguished.  To determine the age distribution, it is helpful to
consider the more fundamental steady-state joint age-size distribution,
$n(x,\t)$, defined as the density of civilizations of size $x$ and age
$\t$, (with $x\leq 2V\t$).  From this joint distribution, the
steady-state size distribution of civilizations of any age is clearly
given by

$$ 
n(x)=\int_{x/2V}^\infty d\t\, n(x,\t),
\eqnai 
$$ 
while the steady-state age distribution of any size
civilizations is given by 
$$ 
n(\t)=\int_0^{2V\t} dx\, n(x,\t). \eqnbi 
$$

By the nature of the warfare process, the joint age-size 
distribution consists of two components, 
$$
n(x,\t)=\CM(x,\t)+{\CI}(\t)\d(x-2V\t). \eqnoi
$$ 
The first term accounts for ``mature'' civilizations which have
experienced at least one war.  The size of such civilizations is
strictly less than the maximum possible size, $x_{\rm max}(\t)=2V\t$, at
a given age $\t$.  The second (singular) term accounts for ``innocent''
civilizations which have not experienced any war during their lifetimes.
The $\d$-function ensures that these innocents are at the maximum size
for a given age.  These two components of the age-size distribution obey
different rate equations.  The equation for the density of innocents is
readily soluble and this facilitates the full solution.

The rate equation for the density of innocent civilizations is
$$
\td{\CI}{\t}=-4V{m_0\over N}{\CI}(\t)+\mu E\d(\t).
\eqnai
$$
The two terms account for the net change of innocent civilizations by
warfare and input, respectively.  Here the fraction of empty space in
the steady state, $E=1/2$, as derived above.  Solving \lasta\ yields
$\CI(\t)={1\over 2}\mu\exp\bigl(-\t\sqrt{8V\mu}\bigr)$, \ie, the
age distribution of innocent civilizations is  purely exponential.

The rate equation for the density of mature civilizations is
$$
\eqalign{
\left(\pd{}{\t}+2V\pd{}{x}\right) \CM(x,\t)=
4Vm_0 \left[\int_0^{\t-x/2V}dy\,{n(y)\over N}\,
{\CM(x+y,\t)\over N}-{\CM(x,\t)\over N}\right] \cr
+4Vm_0{{\CI}(\t)\over N}\,{n(2V\t-x)\over N},\cr}
\eqnbi
$$
where $N=\sqrt{\mu/8V}$ is the steady state civilization density.  The
first term on the right-hand side accounts for the gain of mature
civilizations size $x$ and age $\t$ due to a war between a mature
$(x+y,\t)$-civilization and one (either mature or innocent) of size $y$
and arbitrary age.  Similarly, the second term accounts for the loss of
an $(x,\t)$-civilization due to its undergoing warfare.  The last term
accounts for the creation of $(x,\t)$-civilizations due to an innocent
civilization of size $x$ and age $\t=x/2V$ experiencing its first war
with a civilization of size $2V\t-x$ and arbitrary age.

The full steady state civilization density $n(x)$ and the density of
innocent civilizations $\CI(\t)$ are already known.  Therefore, \lastb\
is a linear integro-differential equation with non-constant
coefficients.  It is possible to reduce \lastb\ to the Klein-Gordon
equation.  This reduction, as well as the solution to the resulting
boundary value problem, is detailed in Appendix B.  From this solution,
we find the steady-state age distribution
$$
n(\t)\propto
\t^{-3/2}\exp\bigl[-(3-\sqrt{8})\t\sqrt{2V\m}\bigr]\eqnoi
$$ 
in the large age limit.  Therefore, the characteristic age is a factor
$(3-\sqrt{8})^{-1}\approx 5.82$ times larger than the characteristic
size divided by the rate of growth $2V$.  Thus a typical domain
survives approximately 6 wars before its ultimate death.  Also, it is
possible to show that the average cluster age
$\av{\t}=\int_0^\infty\int_0^\t d\t\,dx\,\t\,n(x,\t)=2/\sqrt{2V\m}$.
On the other hand, the constant input at unit rate implies that the
domain life expectancy equals $1/\sqrt{2V\m}$.  This inequality
between life expectancy and average age is what may be expected in the
harsh environment defined by our war model.  Domains are especially
vulnerable close to their time of birth, but become progressively more
robust as they grow.  This behavior is akin to that of sea turtles
which are most susceptible to predation immediately after being
hatched.  However, if the hatching survives its initial trek into the
ocean, it has a reasonable chance of living to an old age.

\asubsec{Steady State Properties for Nearly Fair Wars}

In analogy to the case of exactly fair war, we expect that our equation
approach should provide an accurate description for steady-state
properties in the regime of nearly fair war, $\e\ltwid 1$.  The rate
equations in this case are straightforward generalizations of Eqs.~(14)
which, in the steady state, become (taking $V=1/2$)
$$ 
\td{n}{x} = 2m_0\left[\int_0^{x/(1-\e)}
dy\,{n(y)\over N}{n(x+\e y)\over N} -{n(x)\over N}\right]+m_0\delta(x),
\eqnai
$$
$$
\td{m}{x}=
\mu x m(x)-2\mu\int_x^\infty dy\, m(y)+m_0 {m(x)\over N}-{m_0\over N^3}
\int_0^x dz\, m(z)n\left({x-z\over 1+\e}\right)
\int_{x-z\over 1+\e}^\infty dy\,n(y).
\eqnbi
$$

While we have been unable to solve these equations, it is possible to
show that the character of their solution changes as $\e$ decreases from
1 to 0.  This suggests that steady state exists only for a limited range
of $\e>\e_c$, while a different type of solution exists otherwise.  For
this purpose, it is sufficient to consider \lasta.  Using the
transformation
$$
n(x)=m_0\CP(\xi), \qquad \xi={m_0\over N} x,
\eqnoi
$$
one can rewrite the rate equation as
$$
{1\over 2}{d\CP\over d\xi}=
\int_0^{\xi/(1-\e)} d\eta\,\CP(\eta)\CP(\xi+\e \eta)-\CP(\xi),
\eqnoi
$$
which is to be solved subject to the boundary condition $\CP(\xi=0)=1$.
For the case of completely unfair war, $\e=0$, we find
the explicit solution
$$
\CP(\xi)={1\over (1+\xi)^2}.
\eqnoi
$$

This solution has a serious flaw in that the first moment of the
distribution $\CP(\xi)$ is divergent.  However, the first moment of the
unscaled size distribution, $n(x,t)$, is the fraction of covered space
and must clearly be finite.  Thus a physically acceptable steady-state
solution does not exist for $\e=0$.  On the other hand, we have
previously seen that \back1\ does admit a reasonable steady solution,
$\CP(\xi)=e^{-\xi}$, when $\e=1$.  We therefore conclude that the
character of the solution to the rate equation changes for some value of
$\e$ between 0 and 1.  Unfortunately, we are unable to determine the
threshold value of $\e$ below which \back1\ has no physically acceptable
solution.

It is also worth emphasizing that a rate equation description may not
even be applicable when $\e<\e_c$.  When a steady state exists, there is
sufficient empty space available in the system for the steady input to
act as a relatively effective mixing mechanism.  This supports the
notion that a mean-field rate equation approach could provide an exact
description of some steady-state properties, as discussed in the
following section.  On the other hand, if the system coarsens, the
fraction of empty space vanishes and the input becomes progressively
less successful in giving birth to new domains.  Under this
circumstance, it is not evident that the time-dependent rate equations
have the potential to fully capture the time evolution of the system.

\asubsec{Simulation Results in One Dimension}

To test our analytical predictions for $\e=1$ and to map out the
dynamical behavior for general $0\leq\e\leq 1$, we have performed
molecular dynamics simulations for one-dimensional systems with between
1000 to 16000 initial domains, with an input rate of size-less
civilizations per unit length $\m=1/2$, and the growth velocity $V=1/2$.
When a steady state arises, any initial condition would be, in
principle, suitable.  However, to reduce the extent of the early time
regime, we empirically found that a good choice for the initial
condition is a Poisson distribution for both the domain and inter-domain
gap sizes, each with a characteristic length of unity.  A summary of our
numerical results is given in Tables 1 and 2.

For $\e=1$, our analytical predictions for the steady-state values of
the coverage, concentration, and also the form of the size distribution
are confirmed.  In particular, for the size distribution, we find
$n(x)\cong e^{-x/x_0}$, with $x_0$ very close to the exact value of
$1/\sqrt{\mu}=\sqrt{2}$ (Fig.~6).  Simulations for the domain age
distribution also suggest that the asymptotic tail is exponential,
namely $n(\t)\sim e^{-\t/\t_0}$, but with $\t_0=24\pm 2$, which is to be
compared with the analytical result from the rate equation
$\t_0=((3-\sqrt{8})\sqrt{\mu}))^{-1}\approx 8.23$.  The source of this
discrepancy appears to be the existence of significant correlations in
the ages of neighboring civilizations, a feature which would render the
rate equations inaccurate.  This correlation arises because the domain
age remains unaffected by war, so any age correlations which do develop
between neighboring domains persist until all of these domains die.  On
the other hand, the domain size is affected by wars, so that
correlations in domain sizes should be inhibited by the evolution
itself.  To check this hypothesis we measured the size and age
correlation functions, $C_s(r)\equiv\av{s_i s_{i+r}}/\av{s_i}^2-1$ and
$C_\t(r)\equiv\av{\t_i\t_{i+r}}/\av{\t_i}^2-1$, where $s_i$ and $\t_i$
refer to the size and age of the \ith\ domain, respectively.  As
anticipated, $C_s(r)$ is very close to zero for all $r$.  However, the
age correlation function $C_\t(r)$ is systematically negative for $r\leq
5$ (Fig.~7), which implies that old civilizations are less likely to
coexist close to each other.  The absence of size correlations and
presence of age correlations suggests that the mean-field approach
should be quantitatively accurate for the size distribution but not for
the age distribution.

Although the geometrical properties of domains are time independent in
the steady state, their properties as a function of domain age are not
stationary.  This age dependence may provide a useful and deeper
characterization of the steady state.  One such example which appears
especially intriguing is the behavior of the average domain size as a
function of the corresponding domain age (Fig.~8).  This size grows very
slowly and ultimately saturates at a finite value as $t\to\infty$ which
is estimated to be approximately 4.  From the solution to the rate
equations, we find that the average size has the asymptotic form
$\sqrt{2V\over\mu}(2+2\sqrt{2})-{\rm const.}/\t$.  Thus for a domain to
be long-lived, it must not be exceptionally large.

For $\e_c<\e<1$, our simulations show that a steady state is eventually
reached, but that the time needed to attain this steady state grows as
$\e$ approaches $\e_c$ from above (Fig.~9).  Correspondingly, the steady
state values of fraction of empty space, $E$, and the concentration of
domains, $N$, become smaller for decreasing $\e$.  For example, for
$\e=1$ and $\m=1/2$, our analytical results give $E=N=1/2$, while for
$\e=2/3$ and $\m=1/2$, simulations give $E\approx 0.35$ and $N\approx
0.25$.  Additionally, the corresponding domain size and age
distributions for $\e_c<\e<1$ appear to have the same functional forms
as in the fair-war limit of $\e=1$.  In particular, for $\e=2/3$ these
distributions are $n(x)\sim e^{-x/x_0}$, with $x_0=5\pm 0.5$, and
$n(\t)\sim e^{-\t/\t_0}$, with $\t_0=300\pm 50$.  This increase in $x_0$
and $\t_0$ as $\e\to\e_c$ from above has been expected.  As $\e$ is
decreased, conflicts become less devastating for the survivors, so that
they may grow larger and live longer.

When $\e<\e_c$, the system never reaches a steady state (Fig.~9), and
the evolution of a finite size system ends when a single superpower
occupies the entire space.  To quantify this coarsening, we consider
several basic quantities including the average domain size
$\av{x(t)}$, the average maximum domain size $\av{x_{\rm max}(t)}$,
the number density of domains $N(t)$, the fraction of empty space
$E(t)$, the average domain age $\av{\t(t)}$, and the exponents
associated with their asymptotic behavior:

$$
\av{x_{\rm max}(t)} \propto t^{\b(\e)}, \quad
N(t)\propto t^{-\g(\e)}, \quad
E(t)\propto t^{-\g(\e)},\quad
\av{\t(t)}\propto t^{\zeta(\e)}. 
\eqnai
$$

As written, these power laws are found to be non-universal, with
$\e$-dependent exponents (Table~1).  Further, some of these quantities
are interrelated.  For example, $\av{x(t)}=\int dx\,x\,n(x,t)/\int
dx\,n(x,t)=(1-E(t))/N(t)$.  Therefore $\av{x(t)}\propto 1/N(t)\propto
t^{\g(\e)}$.  Furthermore, the fraction of empty space $E(t)$ and
concentration $N(t)$ have the same time dependence.  To establish this,
it is helpful to introduce the normalized domain size distribution
$p(x,t)\equiv n(x,t)/N(t)$.  Numerically, we find that this distribution
approaches a stationary form, $p(x,t)\to p(x)$, in the long time limit,
and with a universal exponent (Table~2),
$$
p(x)\propto x^{-\d} \qquad {\rm when} \quad x\to\infty. 
\eqnbi
$$

This stationarity is, in fact, related to the equivalence between
$N(t)$ and $E(t)$.  Since the concentration of zero size domains is
constant and equal to $\m$ for each inter-domain gap, and since empty
gaps account for a fraction $E(t)$ of the system size, $n(x=0,t)=\m
E(t)$.  On the other hand, the time dependence in $n(x,t)=p(x) N(t)$
appears only through $N(t)$.  We therefore conclude that $N(t)\sim
E(t)$.  It should be noted, however, that this relation is valid only
if $p(x)$ decreases sufficiently fast for large $x$, so that $\int^t
p(x)\, dx$ converges as $t\to \infty$.  Our simulations indicate that
this is indeed the case, namely, $p(x)\sim x^{-\d}$ with $\d\approx
1.7$ (Fig.~10).  It is also possible to relate the distribution
exponent $\d$ with dynamic exponents $\b$ and $\g$ by substituting
\last\ into the relation $1-E(t)=\int^{x_{\rm max}} dx\,xn(x,t)$.
Using $x_{\rm max}(t)\sim t^{\b}$, the integral is found to behave as
$t^{-\g+\b(2-\d)}$, which leads to the exponent relation
$$
\b(2-\d)=\g. \eqnoi
$$
Our simulations agree with \last; \eg, for $\e=0$, we find $\b=1$, in
agreement with the obvious intuition that large domains suffer no damage
and therefore must grow linearly in time.  Correspondingly, we find
$\d\approx 1.7$, and $\g\approx 0.3$.  As the fairness parameter $\e$
increases from 0 to $\e_c$, our simulations indicate that the exponents
$\b$ and $\g$ decrease and both appear to go to zero at $\e=\e_c$.

The borderline case of $\e=\e_c$ can be expected to lead to marginal
behavior which is intermediate to the steady state and coarsening
regimes.  To first establish that $\e_c=0.5$, we examined the evolution
of a specially prepared system in which a single domain is 500 times
larger than all others.  The evolution of this defect domain turns out
to be both useful and computationally efficient way to ascertain whether
the system is in the steady-state or coarsening regimes.  In the steady
state, the defect domain will eventually shrink to the average size.
Thus $\e_c$ can be determined as the point where the defect no longer
shrinks.  This approach gives $0.50\leq\e_c<0.51$ with relatively small
computational effort, suggesting that the value of $\e_c$ equals 1/2.
For $\e=\e_c$, we do find that the average domain age $\av{\t(t)}$ still
grows as a power law in time, $\av{\t(t)}\propto t^{0.61}$, while other
basic observables, such as the fraction of empty space $E(t)$, the
concentration $N(t)$, and the average size $\av{x(t)}$, exhibit
extremely slow variations in time (Fig.~11).  On a double logarithmic
scale, each of these quantities is nearly linear, and a visual fit to
the data suggests an exponent that is approximately 0.1 or less.  A more
careful analysis reveals a weak but systematic curvature in these data,
which suggests that the asymptotic behavior will be slower than a power
law.  However, the time range of the data is insufficient to permit an
unambiguous fit to a logarithmic or other slowly varying time
dependence.

\asubsec{Deterministic Self-Similar Model}

The existence of power-law domain coarsening for $\e<\e_c$ with a simple
relation between length and time scales suggests consideration of a
deterministic self-similar version of our war model to explain the
coarsening exponents.  We discuss here one such example which appears to
be particularly suitable for describing the coarsening dynamics for
$\e=0$.  In this deterministic model, the system starts with a regular
array of domains with spacing $\D x=1$ at $t=0$ (Fig.~12).  The domain
walls move with velocity 1/2 so that the first set of warfare events
takes place at $t=1$.  The outcome is defined to be that every second
domain is annihilated while the remaining domains continue unscathed.
This can be viewed as arising from an infinitesimal difference in the
initial domain sizes.  Also at regular time intervals $\D t=1^+$, new
domains are seeded at the same integer spatial positions of the initial
domains.  Only if the seeding occurs in an empty region does the new
domain grow.  These rules give rise to a pleasing self-similar pattern
of domains which resembles a Sierpinski gasket, except for the filling
of large empty spaces by the continuous input.

For this system, it is straightforward to compute the properties of
domain.  These exhibit strong fluctuations, however, because of the
determinism of the model.  It is therefore convenient to consider
quantities which have been averaged over a finite time range, which we
choose to be between $2^{n-1}$ and $2^n$.  For example, by low-order
enumeration, it is easy to verify that between $t=0$ and $t=2^n$, the
total number of domains in existence over a length $2^n$, starting from
the left edge of figure 12, is given by the sequence 2, 7, 22, 67,
$\dots$ for $n=2$, 3, 4, $\ldots$.  Solving this sequence, the time
integrated density of these domains is asymptotically given by ${5\over
6}({2\over 3})^n$.  If we hypothesize that the density $N(t)$ varies as
$t^{-\g}$, then the corresponding time integrated density is
$$
\int_0^{2^n} N(t)\,dt \sim {t^{1-\g}\over{1-\g}}\Bigm|_0^{2^n}
\propto 2^{n(1-\g)}.
\eqnoi
$$ 
Equating this time integrated density to $(2/3)^n$ as determined above,
we find the exponent $\g=2-\ln 3/\ln 2\approx 0.415$.  In view of the
crudeness of this deterministic model, we regard this exponent value as
being in good agreement with the corresponding numerical result of
$\g\approx 0.33$.

\rchap{SUMMARY AND DISCUSSION}\newcount\nlet\nlet=0

We have introduced an idealized warfare model in which domains grow at a
constant rate and where a contact between two domains results in a war,
with one or both sides suffering casualties.  The long time properties
of the model are fundamentally governed by a fairness parameter $\e$
which quantifies the outcome of a war.  In a war between domains of size
$i$ and $j>i$, the smaller domain is annihilated, while the larger
domain emerges with a size $j-\e i$.  Thus $\e=1$ corresponds to a fair
war where the number of casualties in each domain are equal, while
$\e=0$ corresponds to a completely unfair war where the winner suffers
no casualties.  We have examined the long-time kinetics of this model
for: (a) the heterogeneous process, in an initial distribution of
domains is specified, and after which evolution by domain growth and
intermittent warfare ensues, and (b) the homogeneous process, there is a
continuous input of infinitesimal domains which then undergo growth and
warfare.

For heterogeneous and fair war ($\e=1$), the system naturally coarsens,
with the number of domains decreasing as $1/t$ and their average size
growing as $t$, so that a constant asymptotic coverage arises.  The
domain size distribution obeys scaling in a manner consistent with these
temporal behaviors.  While we have not investigated the extension to
unfair war in detail, the case $\e=1$ clearly provides a lower bound to
the domain size distribution for unfair wars.  Thus it is evident that
the same quantitative linear in time coarsening will occur for both fair
and unfair wars.

In the heterogeneous process, there is a wider range of phenomenology
which is fundamentally controlled by the fairness parameter.  From
simulations and a heuristic argument, there exists a threshold value
$\e_c= 1/2$ which separates a steady state regime, for $\e>\e_c$, from a
regime of continuous coarsening, for $\e<\e_c$.  In the steady state,
the joint age-size distribution of domains provides a comprehensive
characterization.  Interestingly, the domain lifetime is relatively
large, so that a domain typically survives many wars before eventual
death.  In the coarsening regime, the winner of a war suffers relatively
few casualties which promotes the tendency for the oldest clusters in
the system to grow without bound.  This coarsening evolves according to
non-universal $\e$-dependent power laws in time, in which the density of
clusters and the fraction of empty space decay as $t^{-\g(\e)}$, and the
average domain size grows as $t^{\b(\e)}$.  In the completely unfair war
case or $\e=0$, $\b(\e)=1$, as is intuitively clear, but as $\e\to\e_c$
from below $\b(\e)$ appears to vanish.  At the threshold $\e_c$, there
is a very slow evolution of the system which has yet to be understood.

\smallskip
The war model also suggests interesting generalizations; a few
possibilities and some of their attendant consequences are outlined
below.

\medskip\noindent (i) {\bf Size dependent warfare rates}: Suppose that
the process $(i,i+k)\to k$ occurs at a rate $R(i,i+k)$ which has a power
law dependence on the size difference, \ie, $R(i,i+k)=k^\a$.  For fair
war, where the losses of each combatant are equal, the rate equations
for the heterogeneous version of this process are,
$$
\dot c_k(t)=\sum_{i=1}^\infty k^\a c_i(t)c_{i+k}(t)- 
c_k(t)\sum_{i=1}^\infty |k-i|^\a c_i(t)
+\l(c_{k-1}(t)-c_k(t)).
\eqnoi 
$$ 
The constant growth suggests the scaling variable $x=k/\l
t$, but the time-dependent prefactor in the domain size distribution may
be different than $1/t$ (compare with Eq.~(4)).   Making the scaling
ansatz $c_k(t)\propto t^{-a}\CC(x)$ and substituting into \last,
self-consistency obtains only when $a=\a+2$.  Thus the scaling ansatz is
$$
c_k(t)=t^{-(\a+2)}~\CC(x),
\qquad {\rm with}\quad x={k\over \l t}.
\eqnoi
$$
This scaling form predicts
that the size moments behave as power laws in time, $M_n(t)\propto
t^{n-\a-1}$.  Hence, the coverage {\it decays} with time for
positive $\a$, $M_1(t)\sim t^{-\a}$, indicating that aggressiveness
which grows with size disparity leads to extinction (perhaps a lesson
for real civilizations).

We may further generalize to a size-dependent power-law growth rate
$\l=\l_k=k^{-\b}$.  In the peaceful limit of growth and no war, the size
distribution is peaked around $K(t)=[(1+\b)\l t]^{1/(1+\b)}$, while
warfare produces civilizations in the size range $0<k\leq K(t)$.  This
suggests the scaling ansatz
$$
c_k(t)=t^{-{\a+\b+2\over \b+1}}\CC(x), 
\qquad {\rm with}\quad x={k\over K(t)}\leq 1,
\eqnoi 
$$ 
which leads to size moments $M_n(t)\sim t^{(n-\a-\b-1)/(\b+1)}$.
These results are expected to be valid only for $\b>-1$.  For $\b=-1$,
the typical size grows exponentially, while for $\b<-1$ the typical
size diverges in a finite time, \ie, the most aggressive civilization
covers the system.

\medskip\noindent (ii) {\bf Bipolar world}: Consider two mutually
antagonistic species $A$ and $B$ with aggregation occurring when two
same-species civilizations (allies) meet, and war occurring when
dissimilar species meet.  If both species grow at the same constant
rate, the mean-field evolution of $A$-civilizations is described by the
rate equations (in the heterogeneous case)
$$
\eqalign{\dot a_k(t)=\sum_{i=1}^\infty b_i(t)
a_{i+k}(t)-a_k(t)\sum_{j=1}^\infty &
b_j(t) +{1\over 2}\sum_{i=1}^{k-1} a_i(t) a_{k-i}(t)~~~~~~~~~~~~~~~~\cr
-a_k(t)\sum_{j=1}^\infty &a_j(t)
+\l(a_{k-1}(t)-a_k(t)),\cr}
\eqnoi 
$$ 
and similarly for $B$-civilizations.  Here $a_k(t)$ and $b_k(t)$ are the
concentrations of $A$- and $B$-civilizations of size $k$ at time $t$.
The first two terms on the right-hand side account for warfare, the next
two terms account for aggregation, and growth is described by the last
term.

For this process, it is straightforward to determine that the total
number of civilizations $N(t)\simeq 1/t$.  However, the identification
of the appropriate variable in a scaling ansatz for the domain size
distribution is unclear.  In the absence of growth ($\l=0$), a previous
study of the resulting aggregation-annihilation process found that the
typical size scales as $k\sim\sqrt{t}\,\,$\ref{21}.  On the other hand,
growth without aggregation and war leads to a size distribution which is
peaked around $k\sim \l t$, while combined constant growth and
aggregation, but without war\ref{22}, leads to a typical size which
grows as $t\ln t\,$.  A similar ambiguity exists in one dimension since
in single-species aggregation and growth, the typical size grows as
$e^t$\ref{23}.  The homogeneous version of the bipolar world model is
also of interest; a preliminary treatment seems to indicate that a
steady state does not arise even in the case of fair war.

\medskip\noindent (iii) {\bf War in two dimensions}: It is clearly more
realistic to consider our war model in two dimensions, where geometrical
effects naturally play a more prominent role in defining the outcome of
a warfare event.  If one posits that war is a localized event at the
point of contact between two domains, then the continued action of war
will lead to irregularly-shaped domains and possibly to the breakup of
countries.  These are features which are perhaps best investigated by
simulations.

There is a natural simplification which would eliminate the technical
difficulties associated with irregularly-shaped domains, however.
Namely, start with disk-shaped countries and define that after each war
the victor retains a disk shape with its center remaining fixed.  A
further simplification is to consider the situation where all domains
have the same size.  In analogy with the corresponding one-dimensional
system, analytical progress may be possible.  For the equal-size case,
intuition suggests that the coverage will approach a constant value
$M_\infty$ in the long-time limit.  Under the further assumption of
spatial homogeneity, this would suggest that the number density of
civilizations of radius $R$, $N(R)$, will vary as $\sim M_\infty/R^2$.

\bigskip\bigskip\centerline{\bf ACKNOWLEDGMENTS}\smallskip

One of us (S.R.) wishes to thank Glen Swindle for stimulating
discussions and the hospitality of the Aspen Center of Physics, where
this work was initiated.  We also gratefully acknowledge grants from
the ARO and NSF for partial support of this research.

\bigskip\bigskip 

\centerline{\bf APPENDIX A}\medskip 

We wish to solve the rate equation
$$
L{\partial\over \partial L}F(x,L)=F(x,L)+x{\partial\over \partial x}F(x,L)+
\theta(x-3)G(L)\int_1^{x-2}dy F(y,L)F(x-y-1,L),
\eqnoA
$$
that describes the dynamics of equal-size growing domains which mutually
annihilate when they meet.  Here $F(x,L)=L\,n(l,L)/\CN(L)$, where
$n(l,L)dl$ is the number of neighboring civilizations whose centers are
separated by a distance which is between $[l,l+dl]$,
$\CN(L)=\int_L^\infty n(l,L)\,dl$ is the total number of surviving
civilizations, $G(L)=F(x=1,L)$, and $x=l/L$.

To solve \lastA\ we apply the Laplace transform,
$$
\Phi(p,L)=\int_1^\infty dxe^{-px}F(x,L).
\eqnoA
$$
Note that the relation $\CN(L)=\int_L^\infty n(l,L)\,dl$ can be rewritten as
$$
\Phi(0,L)=\int_1^\infty dx F(x,L)=1.
\eqnoA
$$
Combining \backA1\ and \backA2\ gives
$$
\left(L{\partial\over \partial L}+p{\partial\over \partial p}\right)
\Phi(p,L)=-G(L)e^{-p}(1-\Phi^2).
\eqnoA
$$

\lastA\ has been solved previously in the scaling limit of $L\to \infty$, 
where it reduces to the ordinary differential equation\ref{14}
$$
p{d\Phi\over dp}=-G_\infty e^{-p}(1-\Phi^2),
\eqnoA
$$
whose solution is 
$$
\Phi(p)=\tanh\left(G_\infty\int_p^\infty {e^{-q}\over q}dq\right),
\eqnoA
$$
which contains an as yet undetermined numerical factor $G_\infty$.  This
constant is found from a consideration that also establishes the
coverage.  Civilizations cover the same space, $x=1$, in units of
scaled length, so the coverage is clearly
$$ 
M(L)={\int_1^\infty dx F(x,L)\over \int_1^\infty dx xF(x,L)} 
\equiv {1\over \langle x\rangle_L}.  
\eqnoA 
$$ 
Here we use \backA4\ and define $\av{x}_L$ 
by $\av{x}_L=\int_1^\infty dx\, x\,F(x,L)$.

In the long-time limit,  we use the relation
$$
\int_p^\infty {e^{-q}\over q}dq=
- -\ln p -\g -\sum_{n=1}^\infty {(-p)^n\over n\cdot n!},
\eqnoA
$$
to expand $\Phi(p)$ in the small $p$ limit as
$\Phi(p)=1-2\exp(2G_\infty\g)p^{2G_\infty}+\ldots$.  On the other hand,
from the definition of $\Phi(p)$ given in \backA6, we have the
expansion, $\Phi(p)=1-p\langle x\rangle_\infty+\ldots$.  Comparing these
two forms gives the constant $G_\infty=1/2$ and the (scaled) distance
between neighboring civilizations, $\langle
x\rangle_\infty=2\exp(\g)$\ref{14}.  This yields the coverage in the
long-time limit $M_\infty =1/\langle x\rangle_\infty\cong 0.28073$, \ie,
Eq.~(11).  Having established the asymptotic coverage, the number
density $N$ of these equal-size civilizations asymptotically is
$N(t)\sim M_\infty t^{-1}$ in agreement with Eq.~(12).

Now consider the the full time-dependent behavior for which we have to
solve the nonlinear partial differential equation \backA4.  From the
form of the asymptotic solution of the time independent equation, 
it is natural to attempt the ansatz
$$
\Phi(p,L)=\tanh[\Psi(p,L)], \eqnoA
$$ 
which allows us to eliminate the nonlinear  factor $(1-\Phi^2)$.  
Substituting \lastA\ into \backA5\ gives
$$
\left(L{\partial\over \partial L}+p{\partial\over \partial p}\right)
\Psi(p,L)=-G(L)e^{-p}.
\eqnoA
$$
Transforming from the variables $(p,L)$ to $u=\sqrt{pL}$ and
$v=\sqrt{p/L}$ simplifies \lastA\ to
$$
u{\partial\over \partial u}\Psi(u,v)=-G\left({u\over v}\right)e^{-uv}.
\eqnoA
$$
The solution to \lastA\ is now straightforward,
$$
\Psi(u,v)=\int_u^\infty {d\xi\over \xi}~G\left({\xi\over v}\right)
e^{-\xi v}+\chi(v),
\eqnoA
$$
up to an arbitrary function $\chi(v)$.  To determine $\chi(v)$, note
that the definition of $\Phi(p,L)$, (\backA{10}), implies
$\Phi(p,L)\simeq p^{-1}e^{-p}G(L)$ in the large-$p$ limit.  Since $G(L)$
varies over a limited range (as it is clear, \eg, from relation
$G(\infty)=1/2$), we conclude that $\Phi(p,L)\to 0$ and hence,
$\Psi(p,L)\to 0$ as $p\to \infty$.  Choose now $p\sim L\to \infty$; in
the $(u,v)$ variables, this corresponds to $u\to \infty$ and $v$ finite.
Thus the integral in \lastA\ disappears in this limit and we find
$\Psi(\infty,v)=\chi(v)=0$ implying that $\chi(v)$ is trivial.

Returning now to original variables and replacing $\xi$ by $\eta$
defined via $\xi=\eta u$, we rewrite \lastA\ as
$$
\Psi(p,L)=\int_1^\infty {d\eta\over \eta}~G(L\eta)e^{-p\eta}.
\eqnoA
$$
We have thus solved \backA{11}, up to an as yet unknown function $G(L)$.
This function can be found, in principle, from the initial conditions.
Technically, it is convenient to assume that there is a finite
small-size cutoff $L_{\rm min}$ in the initial distribution which we
set to be $L_{\rm min}=1$ without loss of generality.  As an example
initial distribution, consider a shifted Poisson
$$
F(x,1)=\cases {e^{-(x-1)}, \quad &$x\geq 1$, \cr
              0,          \quad &$x<1$. \cr}
\eqnoA
$$
In this case $\Phi(p,1)=(1+p)^{-1}e^{-p}$; therefore,
$G(L)$ is determined from the following equation:
$$
\int_1^\infty {d\eta\over \eta}~G(\eta)e^{-p\eta}=
{\rm Arctanh}[(1+p)^{-1}e^{-p}].
\eqnoA
$$
Although it is impossible to find explicit expression for $G(L)$
in terms of elementary functions, one can readily compute asymptotics
behaviors, \eg, $M_\infty-M(L)\propto L^{-1}$.

Thus the model of equal-size warring civilizations is exactly solvable
in one dimension.  While asymptotic characteristics have been computed
by exploiting previously known results, the complete solution for
arbitrary time is new.  However, several related and interesting
properties have not yet been computed.  One such quantity is the
density of feral space, \ie, the fraction of space that has been
untouched by any civilization.

\bigskip\centerline{\bf  APPENDIX B}\medskip

We outline here a solution to Eqs.~(20) for the joint age-size domain
distribution.  For this purpose, it proves convenient to rescale length
and time by $x \to x\sqrt{2V/\mu}$ and $\t \to \t/\sqrt{2V\mu}$.  In
these rescaled units, the previous results for the size distribution
become
$$
n(x)={e^{-x}\over 2}, \qquad m_0=N={1\over 2}. 
\eqnoB
$$
Then Eq.~(20a) simplifies to ${d{\CI}\over d\t}=-2{\CI}(\t)+
{1\over 2}\d(\t)$, whose solution is
$$
{\CI}(\t)={e^{-2\t}\over 2}.
\eqnoB
$$
This shows that in the steady state regime the number density of
innocent civilizations, $I=\int_0^\infty d\t\,\CI(\t)={1\over
4}$, is one-half of the total number density, $I=N/2$.  Using
Eqs.~(B1)-(B2), we now reduce the rate equation (20b) to
$$
\left(\pd{}{\t}+\pd{}{x}\right)
\CM(x,\t)=2\left[\int_0^{\t-x} dy\,e^{-y}\,
\CM(x+y,\t)-\CM(x,\t)\right]+e^{x-3\t}.
\eqnoB
$$
Introducing $g(x,\t)=e^{3\t-x}\CM(x,\t)$, reduces \lastB\ to
$$
\left(\pd{}{\t}+\pd{}{x}\right)
g(x,\t)=2\int_x^{\t} dy\,g(y,\t)+1.
\eqnoB
$$
Defining now $f(x,\t)\equiv 2\int_x^{\t} dy\,g(y,\t)+1$,
\lastB\ becomes
$$
\left({\partial^2 \over \partial x \partial \t} + 
\p2d{}{x}\right)f=-2f(x,\t).
\eqnoB
$$

One boundary condition, $f(x=\t,\t)=1$, follows directly from the
definition of $f$.  To obtain a second condition, we compare total
number of zero-size civilizations, $n(x=0)$, with the number of
innocent zero-size civilizations, $\CI (x=0)$.  Both quantities
are equal to $1/2$ which means that $\int_0^{\infty} \CM(x=0,\t)
d\t =0$.  Since $\CM(x,\t)$ is nonnegative for all $\t\geq 0$, we
conclude that $\CM(x=0,\t)=0$, which leads to the boundary
condition $\pd{f}{x}|_{x=0}=0$.

\lastB\ simplifies further
after the change of variables, $(x,\t)\to (\a,\b)=(\t,\t-x)$:
$$
{\partial^2 f \over \partial \a \partial \b}=2f(\a,\b).
\eqnoB
$$
This Klein-Gordon equation is to be solved in the region $\a\geq \b\geq
0$, with the boundary conditions 
$$
f|_{\b=0}=1,\qquad{\rm and}\qquad
\pd{f}{\b}\Big |_{\a=\b}=0.
\eqnoB
$$

The symmetry of the governing equation under the exchange of the
variables, $\a \leftrightarrow \b$, suggests seeking a symmetric
solution which depends on a {\it single} variable $\a\b$.  Further
analysis indicates that the variable $z=\sqrt{8\a\b}$ is especially
convenient.  Substituting $f=f(z)$ reduces the Klein-Gordon equation to
the modified Bessel equation,
$$
f''+{1\over z}f'-f=0,
\eqnoB
$$
where the prime denotes differentiation with respect to $z$.  A
potential solution is $f=I_0(z)$.  This satisfies the boundary condition
\backB{2}; however, the boundary condition of \backB{1} is {\it not}
satisfied.  To remove this drawback, we make use of the linearity of the
governing equation and seek a solution of the form

$$
f(\a,\b)=I_0(z)+\b^n g(z).
\eqnoB
$$
The first of the boundary conditions in \backB{3} is manifestly
satisfied (when the index $n$ is positive).  Substituting \lastB\ into
\backB{3} we get

$$
g''+{2n+1\over z}g'-g=0.
\eqnoB
$$
\lastB\ is readily solved to find $g(z)=Az^{-n}I_n(z)$, where $I_n(z)$
is the modified Bessel function of order $n$.  To determine the index $n$
and the amplitude $A$ we substitute \backB{1}, with $g(z)=Az^{-n}I_n(z)$,
into the second boundary condition in \backB{3}.  This yields the relation
$$
\sqrt{2}I_1(z)+2^{1-3n\over 2}AI_{n-1}(z)=0.
\eqnoB
$$
In deriving \lastB\ we used the identities\ref{24}
$$
\eqalign{
I_{n-1}(z)-I_{n+1}(z)&={2n\over z}\,I_n(z), \cr
I_{n-1}(z)+I_{n+1}(z)&=2I'_n(z), \cr}
\eqnoB
$$
and the equality $z=\b\sqrt{8}$ on the diagonal $\a=\b$.  \backB{1}
shows that $n=2$ and $A=-8$.  Thus we determine the desired solution 
to the Klein-Gordon equation with mixed boundary conditions
$$
f(\a,\b)=I_0(z)-{\b\over \a}\,I_2(z).
\eqnoB
$$
Returning to the original variables $(x,\t)$,
and the original joint age-size distribution function $n(x,\t)$,
we get after straightforward computations
$$
n(x,\t)= \CI(x,\t)+\CM(x,\t)=2x\,e^{x-3\t}\,{I_1(z)\over
z}+{e^{-2x}\over 2} \d(x-\t), \qquad \
z\equiv \sqrt{8(\t-x)\t}.
\eqnoB
$$
The sum rule Eq.~(18a) provides a useful self-consistency check.
Substituting \lastB\ into $ n(x)=\int_{x}^\infty d\t\, n(x,\t)$
gives
$$
{e^{-x}\over 2}-{e^{-2x}\over 2}=\int_x^\infty d\t\,
{x\,e^{x-3\t}\over \sqrt{2(\t-x)\t}}\,I_1\left(\sqrt{8(\t-x)\t}\right).
\eqnoB
$$
This identity is indeed satisfied\ref{25}.  Another quantity which can
be calculated exactly is average age, $\av{\t}\equiv \int_0 ^\infty
n(\t) \t d\t$, where $n(\t)=\int_0^{\t} dx\, n(x,\t)$
(Eq.~(18b)). Changing the order of integration and using\ref{25}, we
obtain that $\av{\t}=2$.

Let us finally consider the age distribution of mature civilizations
$\CM(\t)$.  It is given by $\CM(\t)=\int_0^\t dx\CM(x,\t)$.  In the
large-age limit we use the asymptotic relation\ref{24} $I_1(z)\simeq
{e^z\over \sqrt{2\pi z}}$ to estimate the integral.  Thus we arrive at
$$
\CM(\t)\simeq {B\over \t^{3/2}}\, e^{-(3-\sqrt{8})\t}, \qquad
B={(\sqrt{2}+1)^2\over \pi^{1/2}\cdot 2^{7/4}}\cong 0.977629.
\eqnoB
$$
Since the age distribution of innocent civilizations decays as
$e^{-2\t}$, \lastB\ indicates that old civilizations are mostly
mature.  Another interesting computation is the average domain size as a
function of age,
$$
\av{x(\t)}={\int_0^\t dx\,x\,n(x,\t)\over \int_0^\t dx\,n(x,\t)}.
\eqnoB
$$
Using the asymptotic behaviors outlined above, we find
$$
\av{x(\t)}\sim (2+2\sqrt{2})\left(1-{{\rm const.}\over \t}\right).
\eqnoB
$$
\vfill\eject

$$\vbox{\offinterlineskip
\halign{&\vrule#&\strut\ #\ \cr
\multispan{9}\hfill\bf Table 1\hfill\cr
\noalign{\medskip}
\noalign{\hrule}
height6pt&\omit&&\omit&&\omit&&\omit&\cr
&\hfil\qquad$\e$\qquad\hfil&&\hfil\qquad $\b$\qquad\hfil&&\hfil\qquad
$\g$\qquad\hfil&&\hfil
\qquad$\zeta$\qquad\hfil&\cr
height6pt&\omit&&\omit&&\omit&&\omit&\cr
\noalign{\hrule}
height6pt&\omit&&\omit&&\omit&&\omit&\cr
&\hfil $0.45$\hfil&&\hfil $0.46$\hfil&&\hfil $0.15$\hfil&&\hfil
$0.61$\hfil&\cr
height6pt&\omit&&\omit&&\omit&&\omit&\cr
\noalign{\hrule}
height6pt&\omit&&\omit&&\omit&&\omit&\cr
&\hfil $0.4$\hfil&&\hfil $0.55$\hfil&&\hfil $0.18$\hfil&&\hfil
$0.59$\hfil&\cr
height6pt&\omit&&\omit&&\omit&&\omit&\cr
\noalign{\hrule}
height6pt&\omit&&\omit&&\omit&&\omit&\cr
&\hfil $0.2$\hfil&&\hfil $0.77$\hfil&&\hfil $0.25$\hfil&&\hfil
$0.53$\hfil&\cr
height6pt&\omit&&\omit&&\omit&&\omit&\cr
\noalign{\hrule}
height6pt&\omit&&\omit&&\omit&&\omit&\cr
&\hfil $0$\hfil&&\hfil $1$\hfil&&\hfil $0.33$\hfil&&\hfil
$0.35$\hfil&\cr
height6pt&\omit&&\omit&&\omit&&\omit&\cr
\noalign{\hrule}}
}$$
\bigskip

\rtabi The estimated values of the exponents that characterize the time
dependent properties of domains in the coarsening regime, $\e<1/2$.  The
estimated error in these numbers is 5\% or less.
\bigskip

$$\vbox{\offinterlineskip
\halign{&\vrule#&\strut\ #\ \cr
\multispan{7}\hfill\bf Table 2\hfill\cr
\noalign{\medskip}
\noalign{\hrule}
height6pt&\omit&&\omit&&\omit&\cr
&\hfil\qquad $\e$\qquad\hfil&&\hfil\qquad $p(x)$\qquad\hfil&&\hfil\qquad
$p(\t)$\qquad\hfil&\cr
height6pt&\omit&&\omit&&\omit&\cr
\noalign{\hrule}
height6pt&\omit&&\omit&&\omit&\cr
&\hfil $>0.5$\hfil&&\hfil $\rm exponential$\hfil&&\hfil $\rm
exponential$\hfil&\cr
height6pt&\omit&&\omit&&\omit&\cr
\noalign{\hrule}
height6pt&\omit&&\omit&&\omit&\cr
&\hfil $0.5$\hfil&&\hfil $1.75$\hfil&&\hfil $1.47$\hfil&\cr
height6pt&\omit&&\omit&&\omit&\cr
\noalign{\hrule}
height6pt&\omit&&\omit&&\omit&\cr
&\hfil $0.45$\hfil&&\hfil $1.79$\hfil&&\hfil $1.49$\hfil&\cr
height6pt&\omit&&\omit&&\omit&\cr
\noalign{\hrule}
height6pt&\omit&&\omit&&\omit&\cr
&\hfil $0.4$\hfil&&\hfil $1.72$\hfil&&\hfil $1.51$\hfil&\cr
height6pt&\omit&&\omit&&\omit&\cr
\noalign{\hrule}
height6pt&\omit&&\omit&&\omit&\cr
&\hfil $0.2$\hfil&&\hfil $1.66$\hfil&&\hfil $1.63$\hfil&\cr
height6pt&\omit&&\omit&&\omit&\cr
\noalign{\hrule}
height6pt&\omit&&\omit&&\omit&\cr
&\hfil $0$\hfil&&\hfil $1.67$\hfil&&\hfil $1.67$\hfil&\cr
height6pt&\omit&&\omit&&\omit&\cr
\noalign{\hrule}
}}$$
\bigskip

\rtabi The characteristic exponents of the age and size
distributions. The estimated error in these numbers is 10\% or less.

\bigskip\bigskip
\vfill\eject
\centerline{\bf REFERENCES}\medskip

\refi  J.~D.~Gunton, M.~San Miguel, and P.~S.~Sahni, in {\sl Phase
       Transitions and Critical Phenomena}, eds.\ C.~Domb and 
       J.~L.~Lebowitz (Academic Press, London, 1983), vol.\ 8.  

\refi  J.~S.~Langer, in {\sl Solids Far From Equilibrium}, ed.\
       C.~Godr\`eche (Cambridge University Press, Cambridge, 1992).

\refi  A comprehensive recent review of the coarsening dynamics
       is given by A.~J.~Bray, \ap 43 357 1994 .

\refi  D.~A.~Beysens, and C.~M.~Knobler, \prl 57 1433 1986 ; J.~L.~Viovy,
       D.~A.~Beysens, and C.~M.~Knobler, \pra 37 4965 1988 ; 
       D. Fritter, C. M. Knobler, and D.~A.~Beysens, \pra 43 2858 1991 . 

\refi  P. Meakin, \rpp 55 157 1992 .

\refi  K.~Kawasaki, in {\sl Phase Transitions and Critical Phenomena},
       eds.\  C.~Domb and M. S. Green (Academic Press, London, 1972), 
       vol.\ 2. 

\refi  D.~Weaire and N.~Rivier, \contp 25 59 1984 ; J.~A.~Glazier and
       J.~Stavans, \pra 40 7398 1989 .

\refi  J. Stavans, \rpp 56 733 1993 .

\refi  B.~Derrida, C.~Godreche, and I.~Yekutieli, 
       \pra 44 6241 1991 .

\refi  T.~Nagai and K.~Kawasaki, \pA 120 587 1983 ;
       K.~Kawasaki and T.~Nagai, \pA 121 175 1983 .

\refi  J.~Zhuo, G.~Murthy, and S.~Redner, \jpa 25 5889 1992 .

\refi  J.~Carr and R.~Pego, \prsl 436 569 1992 .

\refi  S.~N.~Majumdar and D.~A.~Huse, \pre 52 270 1995 .

\refi  T.~Nagai and K.~Kawasaki, \pA 134 483 1986 ;
       K.~Kawasaki, A.~Ogawa, and T.~Nagai, \pB 149 97 1988 .

\refi  A.~D.~Rutenberg and A.~J.~Bray, \pre 50 1900 1994 .

\refi  A.~J.~Bray, B.~Derrida, and C.~Godreche, \eul 27 175 1994 .

\refi  A.~J.~Bray and B.~Derrida, \pre 51 1633 1995 .

\refi  See \eg, J.~Keegan, {\sl A History of Warfare}, (Alfred A. Knopf,
       New York, 1993); L.~N.~Gumilev, {\sl Searches for an Imaginary
       Kingdom: The Legend of the Kingdom of Prester John}, (Cambridge
       University Press, New York, 1987).

\refi  The rate equations and their solution for this competitive process
       in the absence cluster growth was given in S. Redner, D. ben-Avraham,
       and B. Kahng, \jpa 20 1231 1987 .

\refi  A similar algorithm was developed for the simulation of ballistic
       annihilation processes in P.~L.~Krapivsky, S.~Redner, and F.~Leyvraz,
       \pre 51 3977 1995 .

\refi  P.~L.~Krapivsky, \pA 198 135 1993 ; E. Ben-Naim and
       P.~L.~Krapivsky, \pre 52 6066 1995 .

\refi  P. L. Krapivsky and S. Redner, unpublished. 

\refi  K.~Sekimoto,  {\sl Int.~J.~Mod.~Phys.~B} {\bf 5}, 1843 (1991).

\refi  C.~M.~Bender and S.~A.~Orszag, {\sl Advanced Mathematical Methods
       for Scientists and Engineers} (McGraw-Hill, New York, 1978).

\refi  A.~P.~Prudnikov, Yu.~A.~Brychkov, O.~I.~Marichev, {\sl Integrals
       and Series} (Gordon and Breach Science Publishers, New York,
       1986), v.2, p.309.
\vfill\eject
\psfig{figure=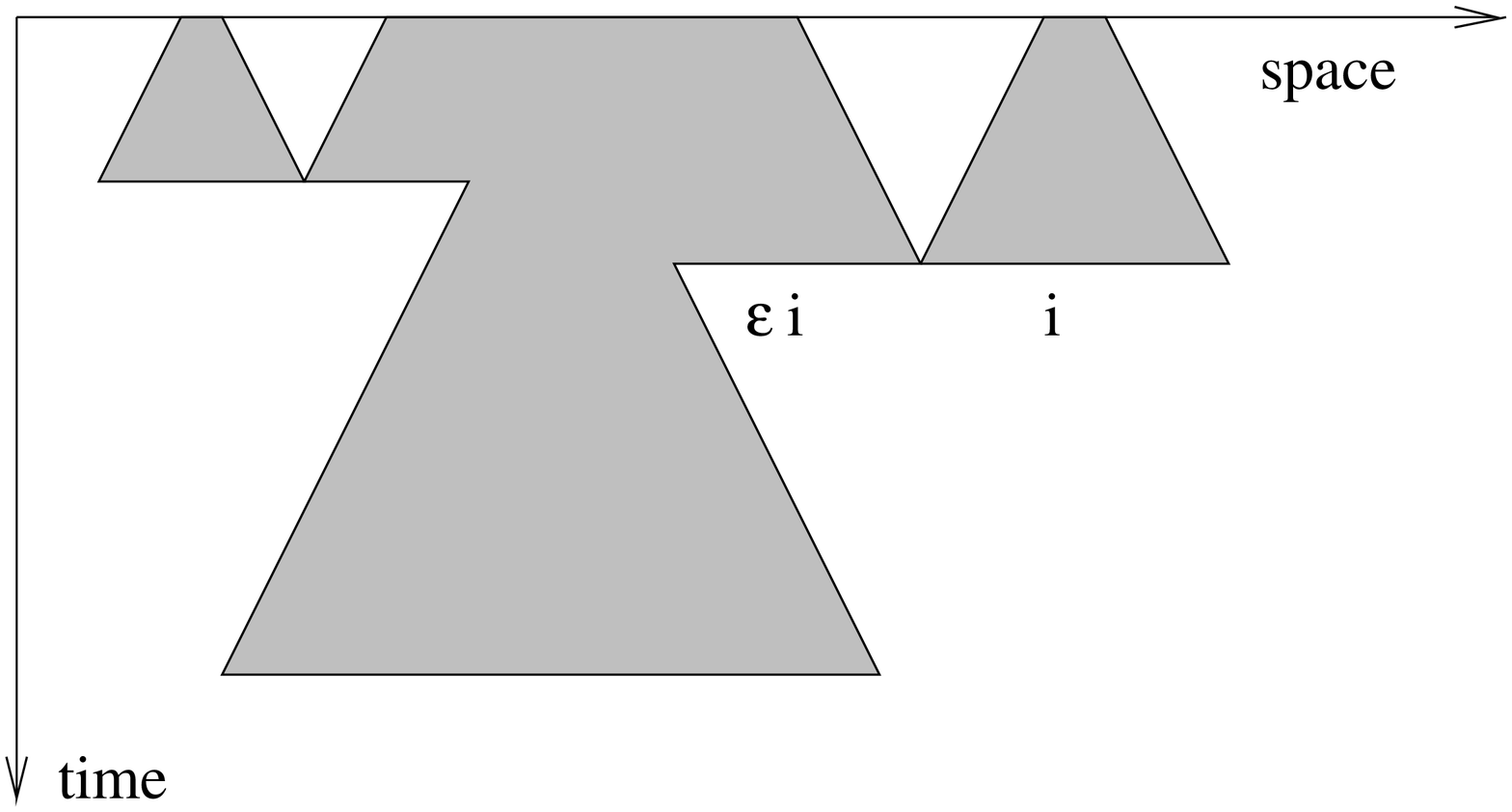,height=4in,width=5in}
\vskip 1 in\bigskip\bigskip
\rfigi Space-time dynamics of vicious civilizations in one dimension.
Shown is the heterogeneous version of the war model where the system
evolves from a fixed initial state.  When two civilizations of size $i$
and $j>i$ meet, the result is a diminished civilization of size
$j-\e\,i$.  By definition, casualties occur at the point of contact, so
that the side of the large civilization on the battle front retreats by
$\e\,i$ while the other side is not affected.
\vfill\eject
\psfig{figure=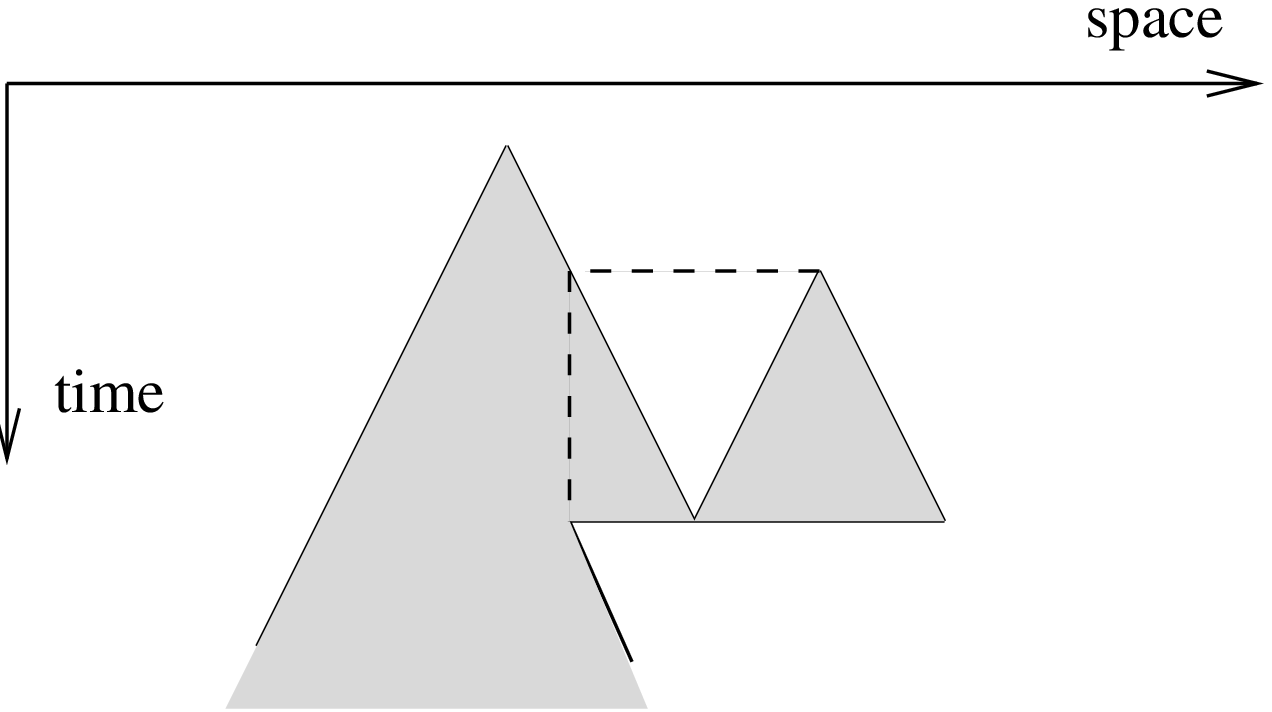,height=4in,width=5in}
\vskip 1 in\bigskip\bigskip
\rfigi Space-time evolution of two domains for the special case of
$\e=1/2$.  The position of the frontier of the winner immediately after
the war and at the birth-time of the smaller combatant is the same.
\vfill\eject
\psfig{figure=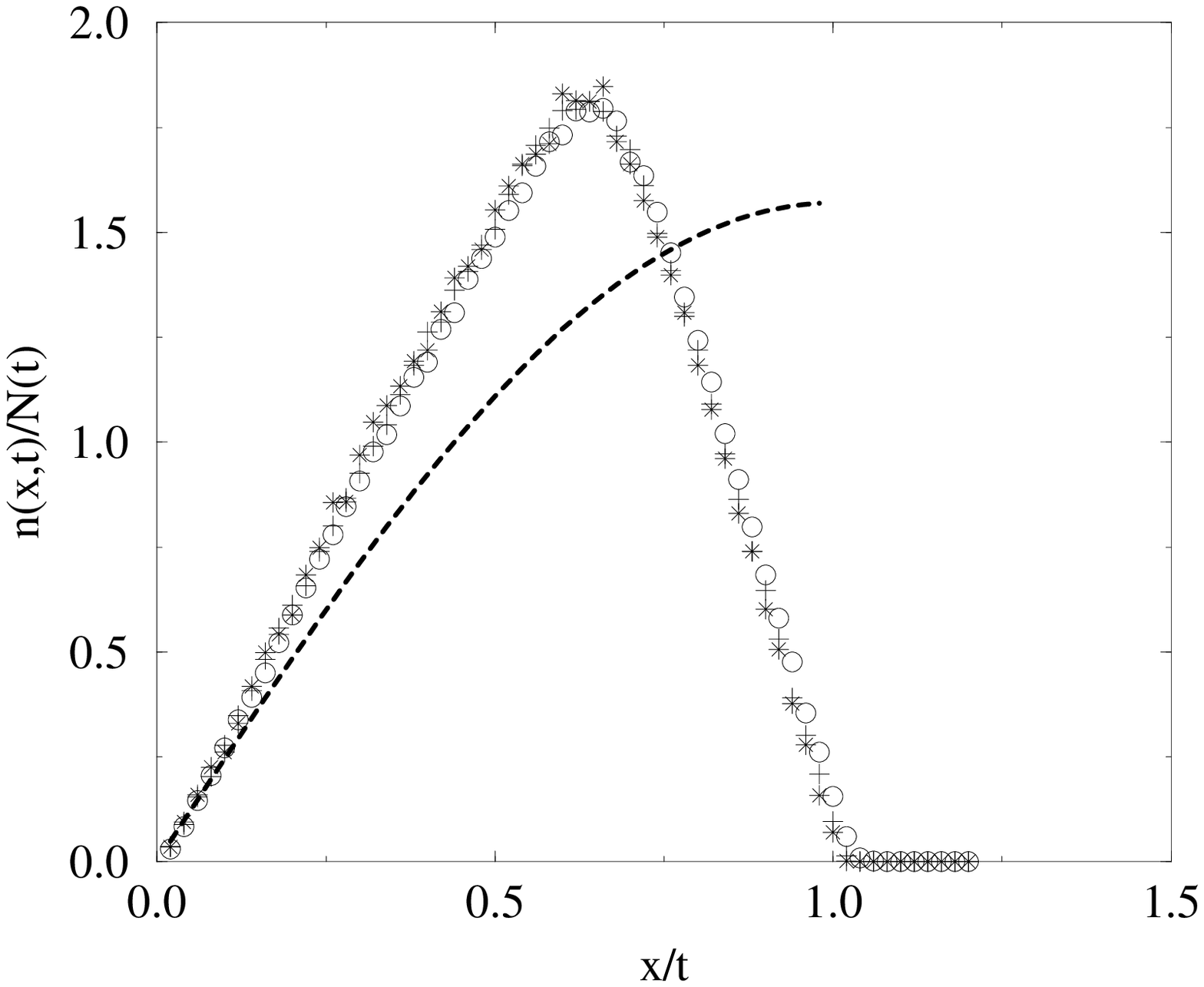,height=6in,width=5in}
\vskip 1 in\bigskip\bigskip
\rfigi Scaled civilization size distribution, $n(x,t)/N(t)$, from
numerical simulations of 500 configurations of heterogeneous war with
$10^5$ initial domains in one dimension.  The data are for $t\approx
1.5^{10}\approx 57.7$ ($\circ$), $t\approx 1.5^{12}\approx 129.7$
($+$), and $t\approx 1.5^{14}\approx 291.9$ ($\ast$).  Also shown is
the corresponding mean-field result ${\pi\over 2}\sin{\pi x\over 2t}$
(dashed).
\vfill\eject
\psfig{figure=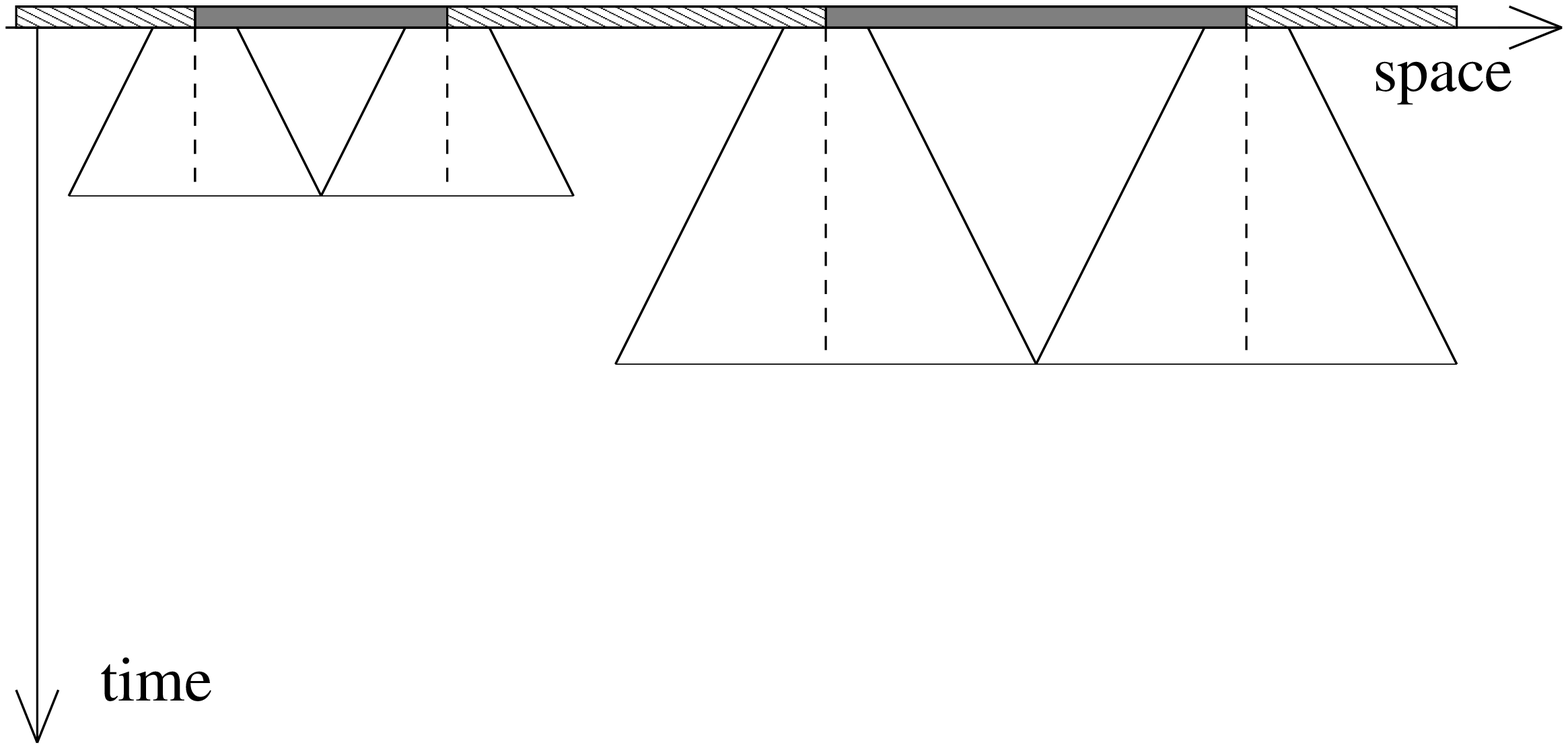,height=4in,width=5in}
\vskip 1 in\bigskip\bigskip
\rfigi Schematic illustration of the equivalence between the space-time
evolution in the one-dimensional war model with equal size domains, and
the ``dual'' problem of coarsening of domains through the successive
elimination of the smallest domain.  In this equivalence, the centers of
the domains in the war model are identified with the positions of domain
walls in the coarsening process (dashed lines).  The two domains which
are eliminated in the dual problem are indicated by the dark shading.
\vfill\eject
\psfig{figure=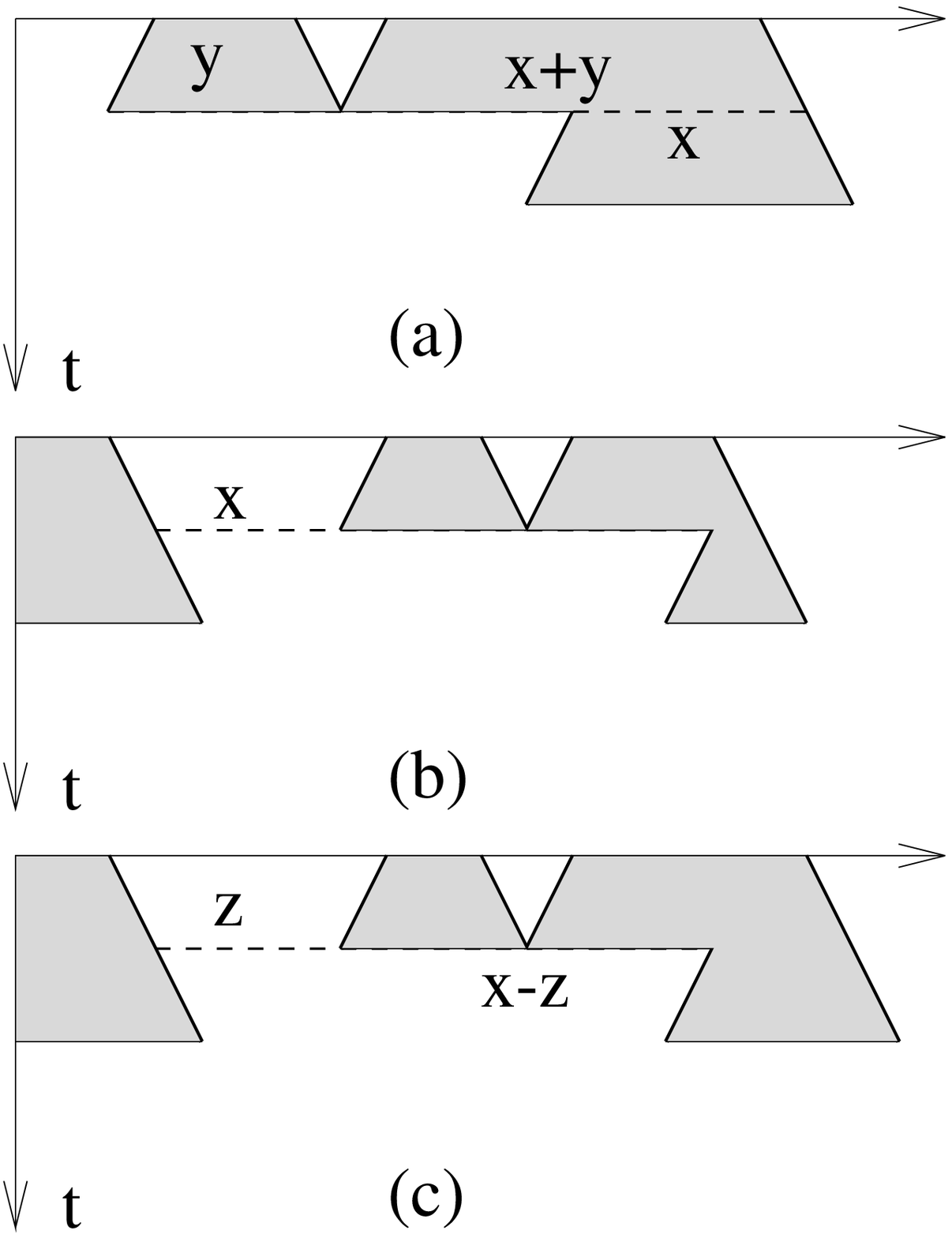,height=6in,width=5in}
\vskip 1 in\bigskip\bigskip
\rfigi Correspondence between various interaction events and the terms
in the rate equation, Eqs.~(14). (a) Illustration of the first term on
the right-hand side of Eq.~(14a), (b) the third and (c) the fourth terms
on the right-hand side of Eq.~(14b).
\vfill\eject
\psfig{figure=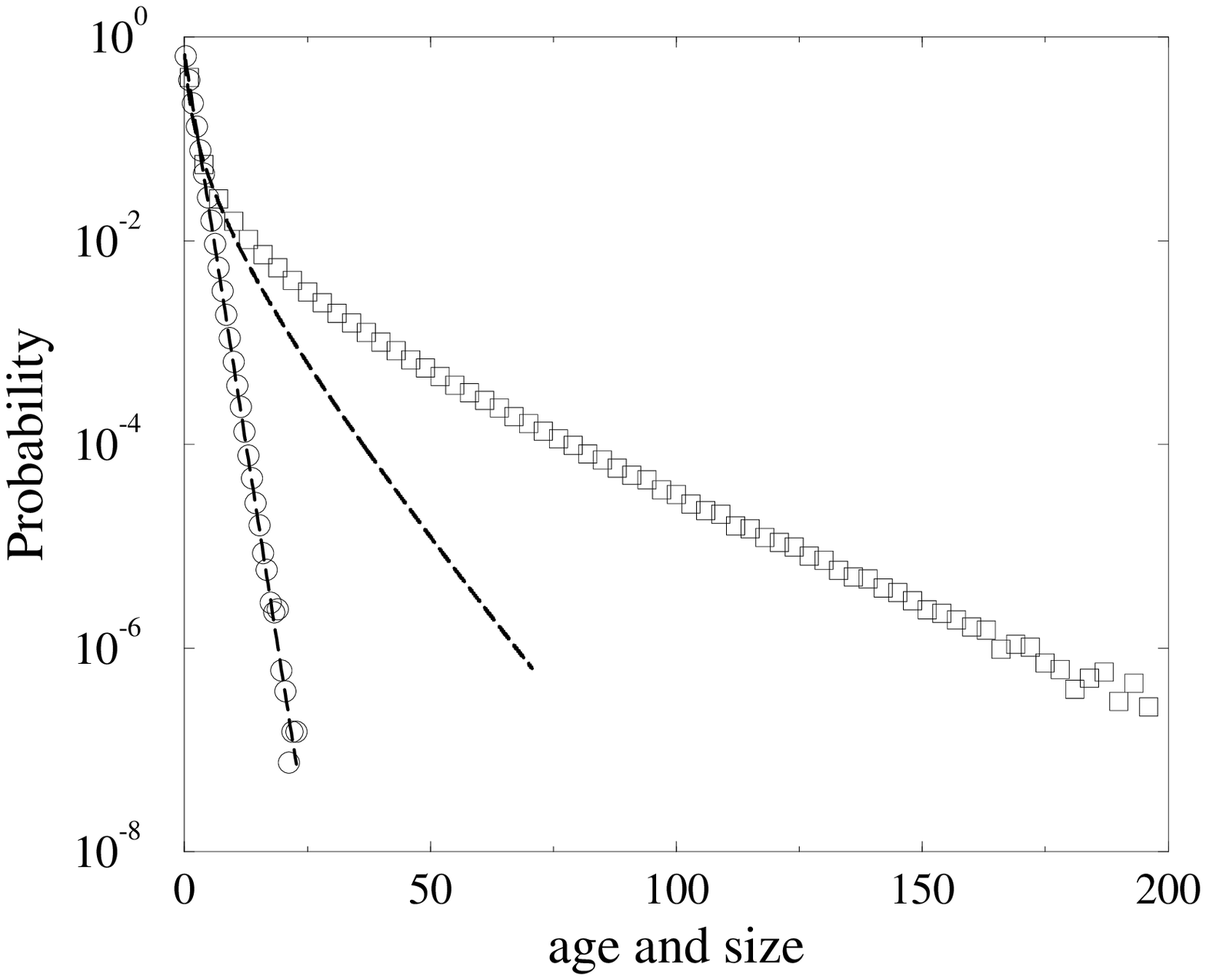,height=6in,width=5in}
\vskip 1 in\bigskip\bigskip
\rfigi Steady state domain size ($\circ$) and age ($\square$)
distributions for the case $\e=1$.  The corresponding predictions from
the solution to the rate equations are also shown (dashed lines).
\vfill\eject
\psfig{figure=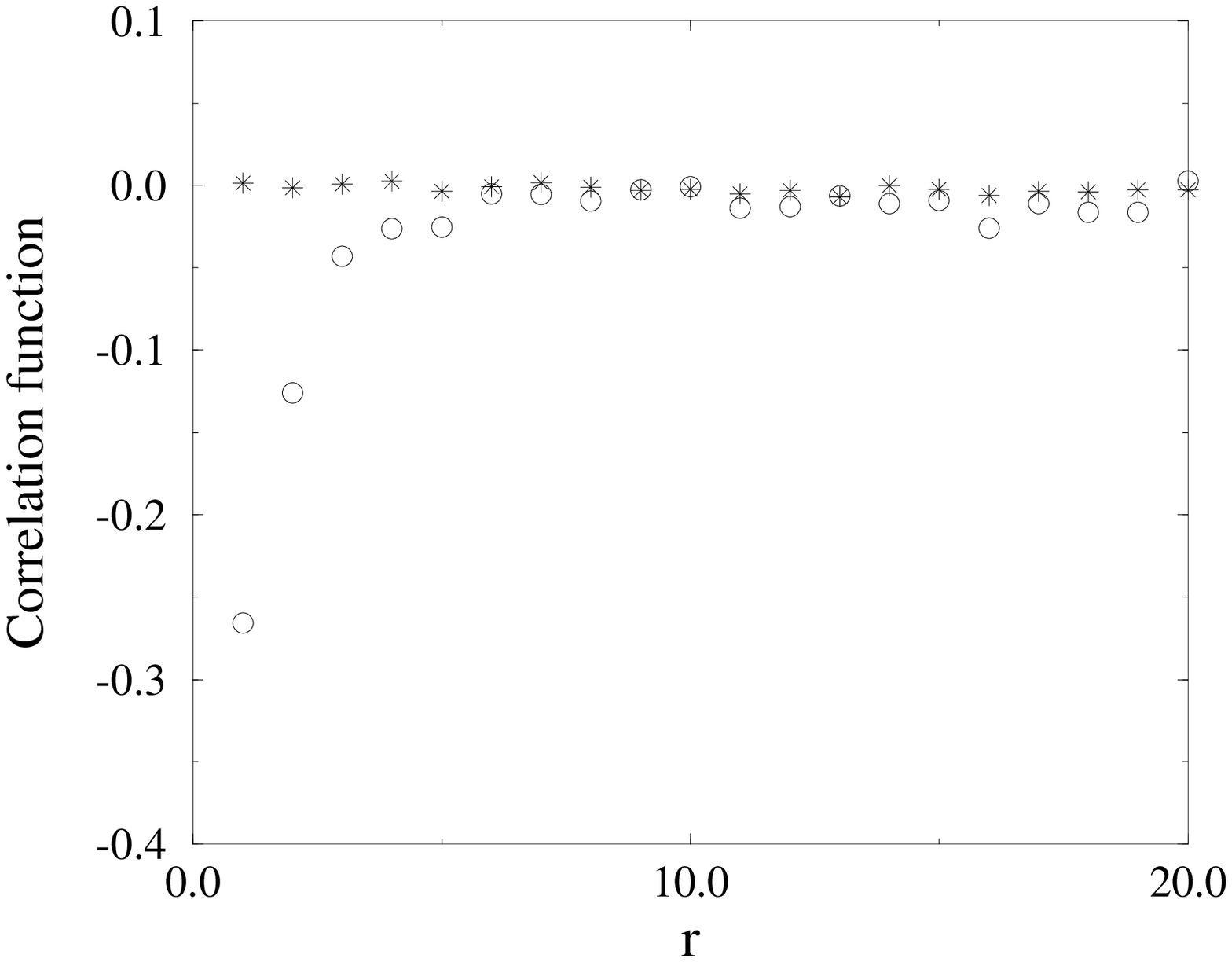,height=6in,width=5in}
\vskip 1 in\bigskip\bigskip
\rfigi The $r$-dependence of the steady-state size ($\ast$) and age
($\circ$) correlation functions, $C_s(r)\equiv\av{s_i
s_{i+r}}/\av{s_i}^2-1$ and
$C_\t(r)\equiv\av{\t_i\t_{i+r}}/\av{\t_i}^2-1$, respectively, for the
case of fair war $\e=1$.  
vfill\eject
\psfig{figure=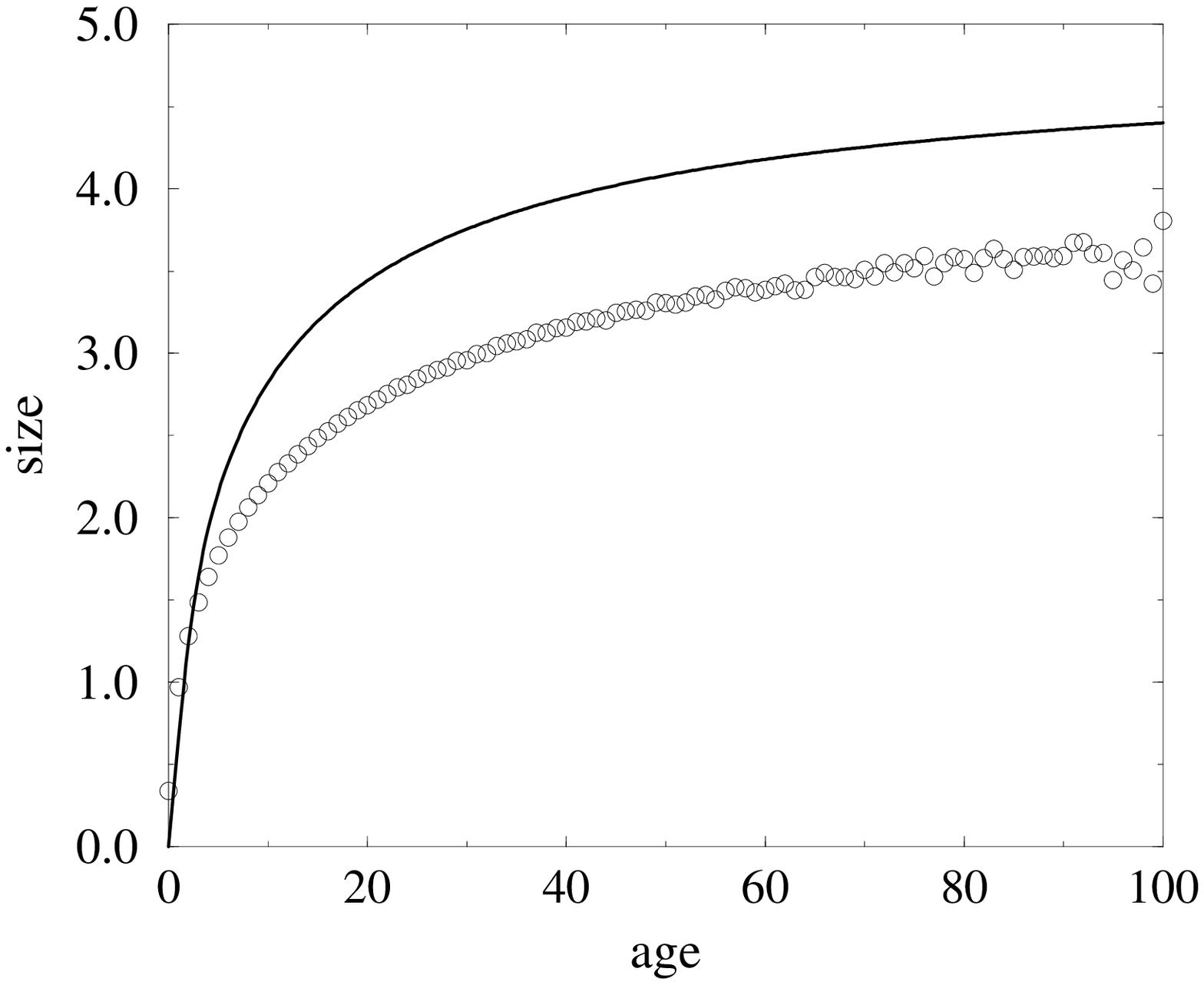,height=6in,width=5in}
\vskip 1 in\bigskip\bigskip
\rfigi Average domain size as a function of domain age in the steady
state for the fair war case of $\e=1$.  Shown is are the simulation
results ($\square$) and the predictions based on the solution to the
rate equations (solid line).
\vfill\eject
\psfig{figure=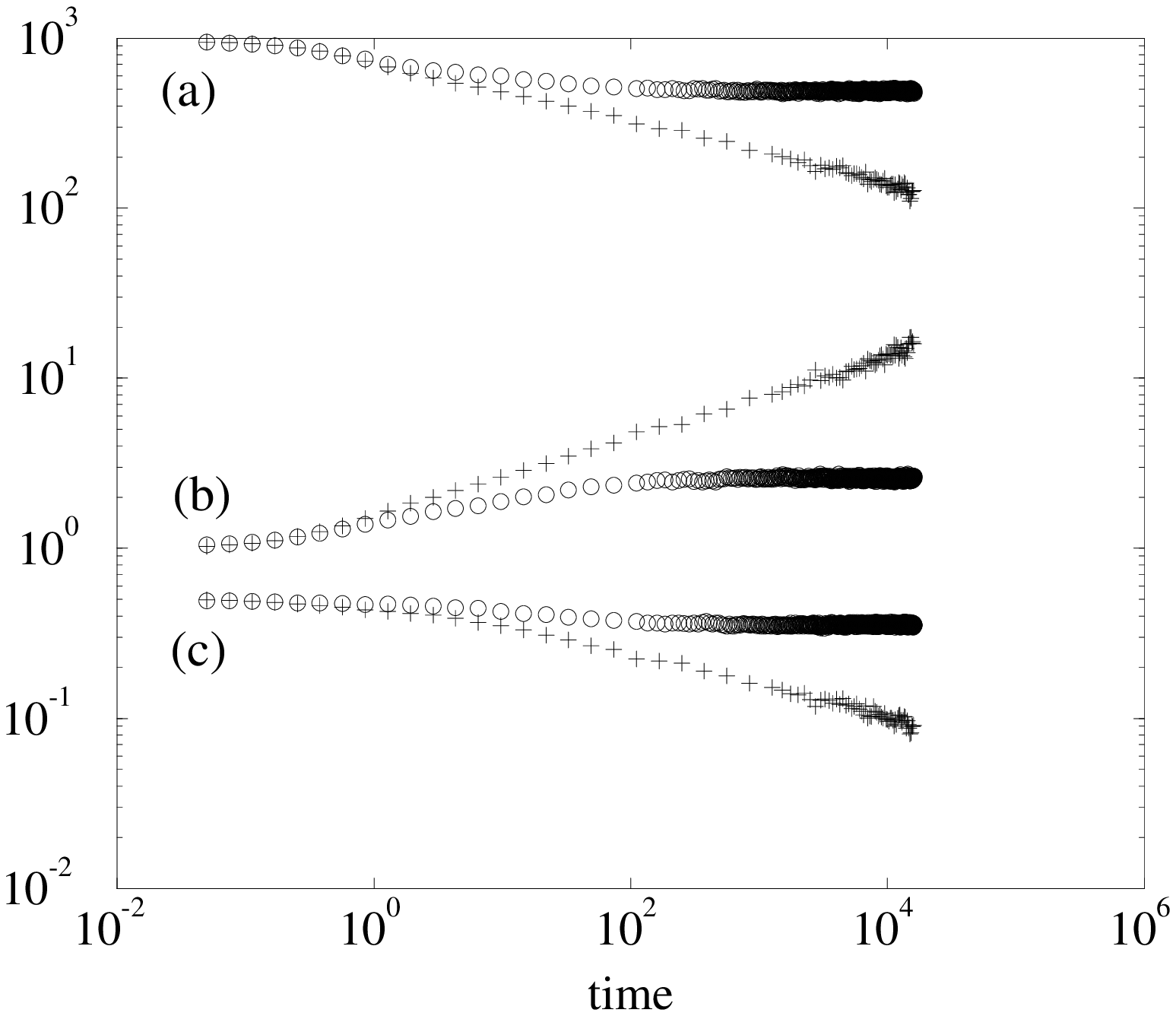,height=6in,width=5in}
\vskip 1 in\bigskip\bigskip
\rfigi Representative simulation results for time-dependent quantities
for a system that has reached a steady state, $\e=2/3$ $(\circ)$, and
for a system which perpetually coarsens, $\e=2/5$ ($+$).  Shown are:
(a) the total number of domains, (b) the average domain size, (c) the
fraction of empty space, and (d) the average domain age.  This data is
based on 20 configurations of system with $10^3$ domains initially.
\vfill\eject
\psfig{figure=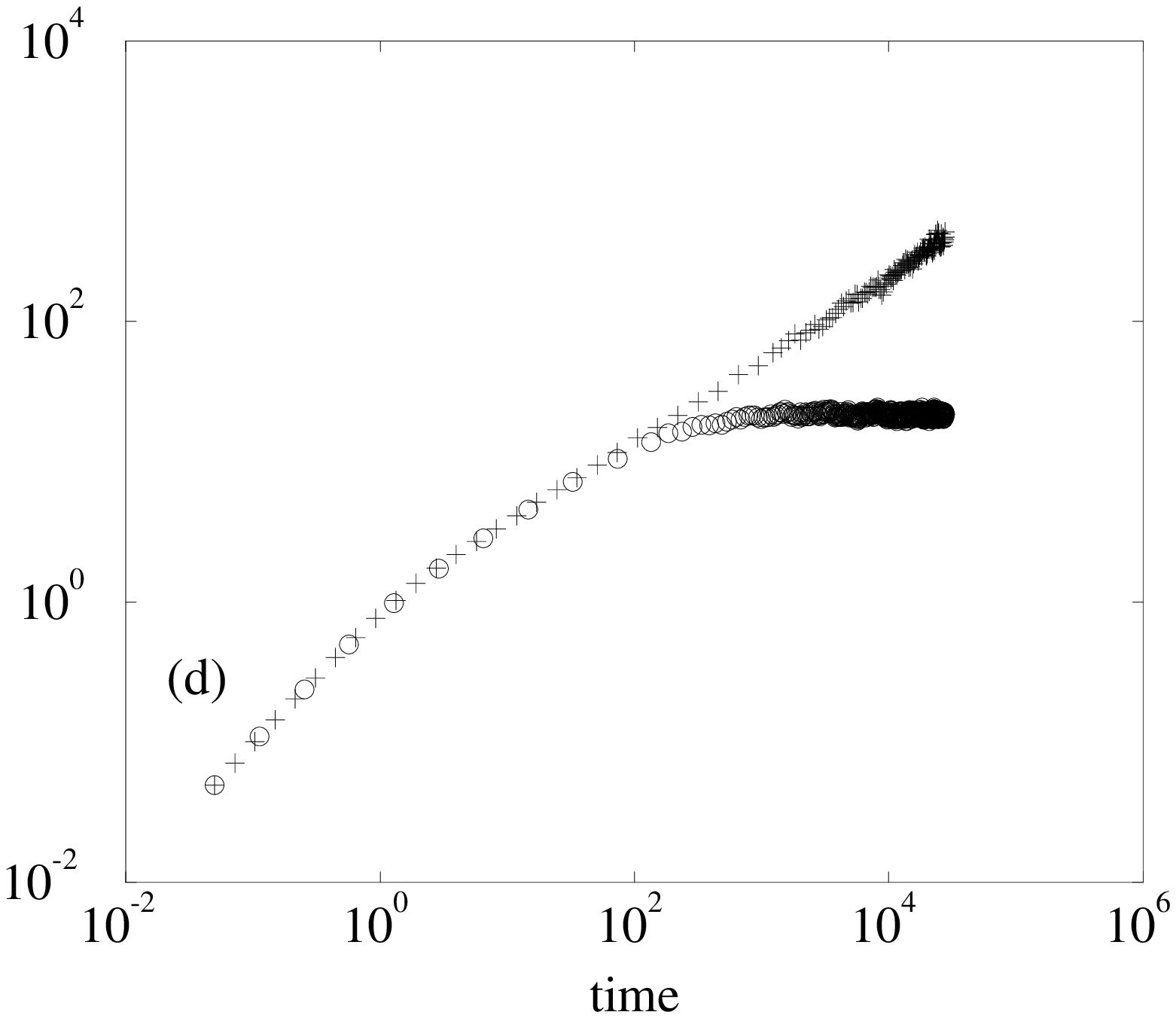,height=6in,width=5in}
\bigskip
\centerline{Figure 9 (d)}
\vfill\eject
\psfig{figure=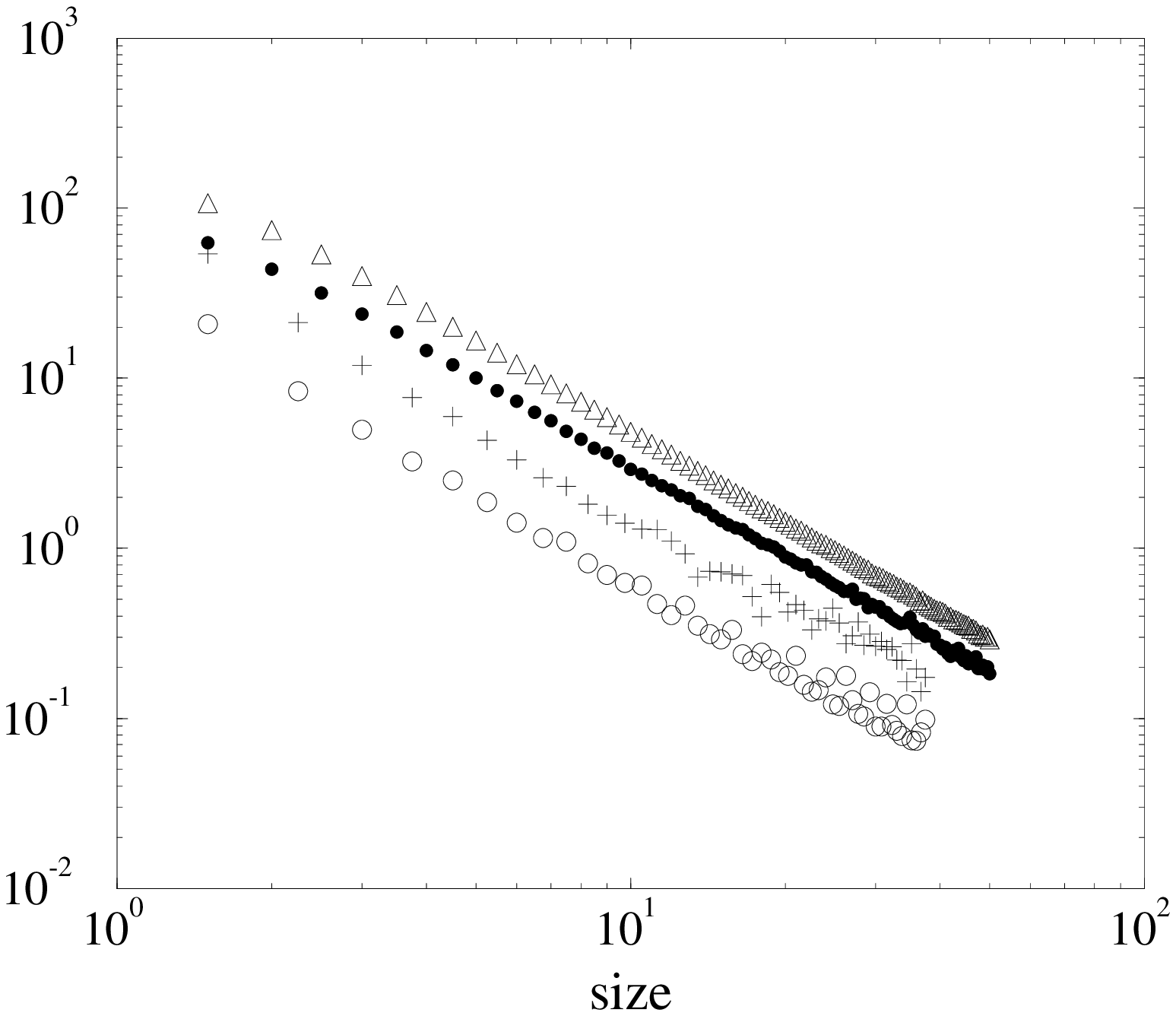,height=6in,width=5in}
\vskip 1 in\bigskip\bigskip
\rfigi Domain size distribution in the coarsening regime for $\e=0$
($\circ$), $\e=0.2$ ($+$), $\e=0.4$ ($\bullet$), and $\e=1/2$
{$\Delta$).  To distinguish the different data sets, the points for each
value of $\e$ have been shifted vertically by a small fixed amount.

\psfig{figure=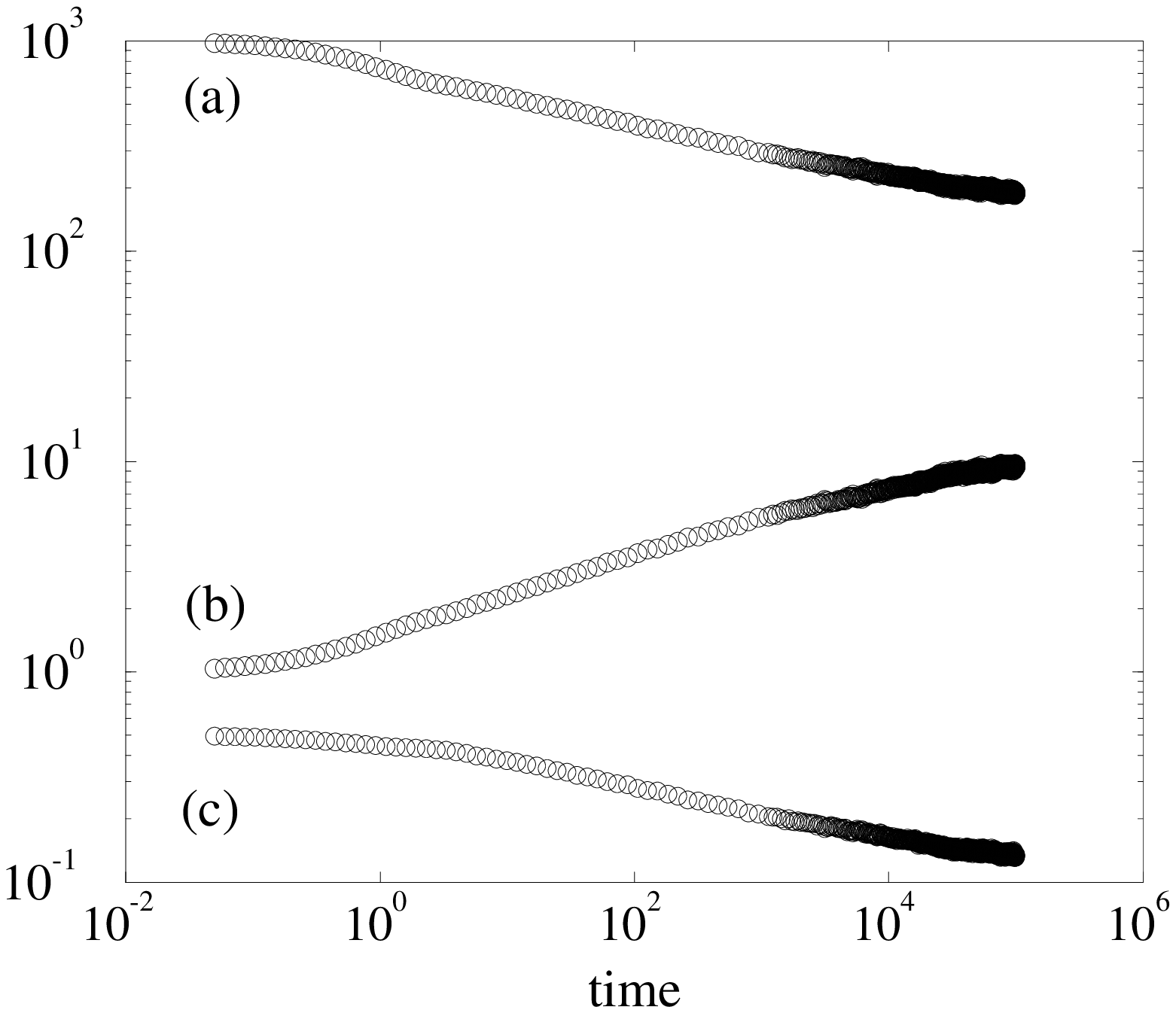,height=6in,width=5in}
\vskip 1 in\bigskip\bigskip
\rfigi Simulation results for time-dependent quantities for the marginal
case of $\e=1/2$ Shown are: (a) the total number of domains, (b) the
average domain size, (c) the fraction of empty space, and (d) the
average domain age.  This data is based on 100 configurations of system
with $10^3$ domains initially.  
\vfill\eject
\psfig{figure=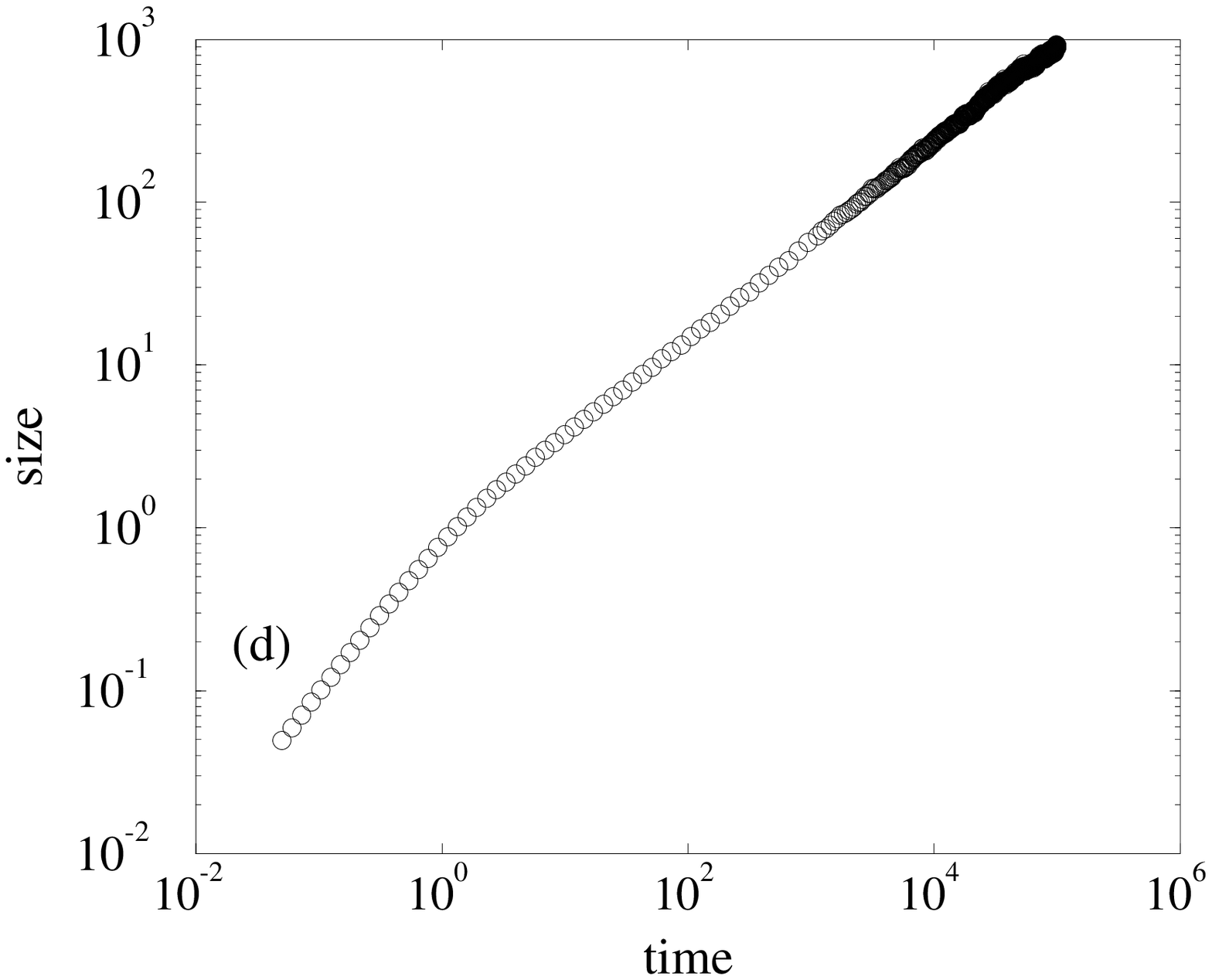,height=6in,width=5in}
\bigskip
\centerline{Figure 11 (d)}
\vfill\eject
\psfig{figure=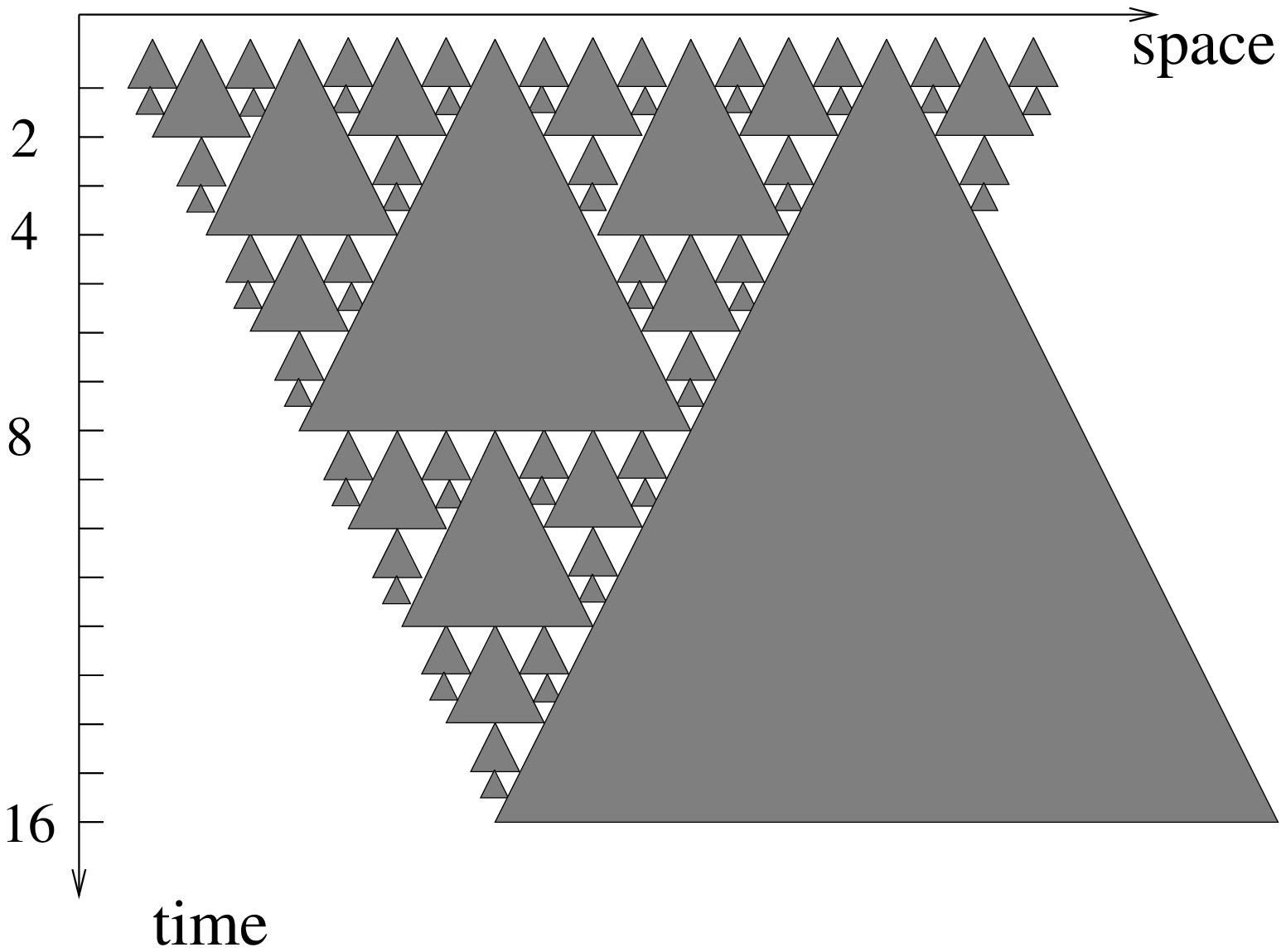,height=5in,width=5in}
\vskip 1 in\bigskip\bigskip

\rfigi Illustration of the deterministic version of war dynamics for the
case $\e=0$.  Domains are created at regular lattice points and at
regular unit time intervals.  When equal size domains meet, a
deterministic rule is implement to decide the survivor. 
               
\vfill\eject\bye

%% file: macros.tex
\def\gl{\mathrel{\raise1ex\hbox{$>$\kern-.75em\lower1ex\hbox{$<$}}}}
\def\lg{\mathrel{\raise1ex\hbox{$<$\kern-.75em\lower1ex\hbox{$>$}}}}
\def\gtwid{\mathrel{\raise.3ex\hbox{$>$\kern-.75em\lower1ex\hbox{$\sim$}}}}
\def\ltwid{\mathrel{\raise.3ex\hbox{$<$\kern-.75em\lower1ex\hbox{$\sim$}}}}
\def\sqr#1#2{{\vcenter{\hrule height.#2pt
      \hbox{\vrule width.#2pt height#1pt \kern#1pt
         \vrule width.#2pt}
      \hrule height.#2pt}}}
\def\square{\mathchoice\sqr34\sqr34\sqr{2.1}3\sqr{1.5}3}

\overfullrule=0pt


\def\eg{\hbox{{\it e.\ g.}}}\def\ie{\hbox{{\it i.\ e.}}}\def\etc{{\it etc.}}

\def\apr{{\it a priori}}


\def\leaderfill{\leaders\hbox to 1em{\hss.\hss}\hfill}


\def\CC{\hbox{{$\cal C$}}}

\def\CI{\hbox{{$\cal I$}}}

\def\CM{\hbox{{$\cal M$}}} 
\def\CN{\hbox{{$\cal N$}}} 
 
\def\CP{\hbox{{$\cal P$}}}

\def\ref#1{~[#1]}
\newcount\eqnum \eqnum=0  
\newcount\eqnA\eqnA=0\newcount\eqnB\eqnB=0\newcount\eqnC\eqnC=0\newcount\eqnD\eqnD=0
\def\eqnoi{\global\advance\eqnum by 1\eqno(\the\eqnum)}
\def\eqnai{\global\advance\eqnum by 1\eqno(\the\eqnum{a})}
\def\eqnbi{\eqno(\the\eqnum{b})}
\def\eqnci{\eqno(\the\eqnum{c})}
\def\eqnoA{\global\advance\eqnA by 1\eqno({\rm A}\the\eqnA)}
\def\eqnoB{\global\advance\eqnB by 1\eqno({\rm B}\the\eqnB)}
\def\eqnoC{\global\advance\eqnC by 1\eqno({\rm C}\the\eqnC)}
\def\eqnoD{\global\advance\eqnD by 1\eqno({\rm D}\the\eqnD)}
\def\back#1{{\advance\eqnum by-#1 Eq.~(\the\eqnum)}}
\def\backa#1{{\advance\eqnum by-#1 Eq.~(\the\eqnum(a))}}
\def\backb#1{{\advance\eqnum by-#1 Eq.~(\the\eqnum(b))}}
\def\backc#1{{\advance\eqnum by-#1 Eq.~(\the\eqnum(c))}}
\def\backs#1{{\advance\eqnum by-#1 Eqs.~(\the\eqnum)}}
\def\backn#1{{\advance\eqnum by-#1 (\the\eqnum)}}
\def\backA#1{{\advance\eqnA by-#1 Eq.~(A\the\eqnA)}}
\def\backB#1{{\advance\eqnB by-#1 Eq.~(B\the\eqnB)}}
\def\backC#1{{\advance\eqnC by-#1 Eq.~(C\the\eqnC)}}
\def\backD#1{{\advance\eqnD by-#1 Eq.~(D\the\eqnD)}}
\def\last{{Eq.~(\the\eqnum)}}                     
\def\lasta{{Eq.~(\the\eqnum{a})}}                 
\def\lastb{{Eq.~(\the\eqnum{b})}}                 
\def\lastc{{Eq.~(\the\eqnum{c})}}                 
\def\lasts{{Eqs.~(\the\eqnum)}}                   
\def\lastn{{(\the\eqnum)}}                      
\def\lastA{{Eq.~(A\the\eqnA)}}\def\lastB{{Eq.~(B\the\eqnB)}}
\def\lastC{{Eq.~(C\the\eqnC)}}\def\lastD{{Eq.~(D\the\eqnD)}}
\newcount\refnum\refnum=0  
\def\refi{\smallskip\global\advance\refnum by 1\item{\the\refnum.}}

\newcount\rfignum\rfignum=0  
\def\rfigi{\medskip\global\advance\rfignum by 1\item{Figure \the\rfignum.}}

\newcount\fignum\fignum=0  
\def\figi{\global\advance\fignum by 1 Fig.~\the\fignum}

\newcount\rtabnum\rtabnum=0  
\def\rtabi{\medskip\global\advance\rtabnum by 1\item{Table \the\rtabnum.}}

\newcount\tabnum\tabnum=0  
\def\tabi{\global\advance\tabnum by 1 Table~\the\tabnum}

\newcount\secnum\secnum=0 
\def\chap#1{\global\advance\secnum by 1
\bigskip\centerline{\bf{\the\secnum}. #1}\smallskip\noindent}

\def\lsubsec#1{\global\advance\secnum by 1
\smallskip{\bf\noindent{\the\secnum. #1}}}
\def\tlsubsec#1{\global\advance\secnum by 1
{\bf\noindent{\the\secnum. #1}}}

\newcount\nlet\nlet=0
\def\numblet{\relax \global\advance\nlet by 1
\ifcase \nlet \ \or a\or b\or c\or d\or e\or
f\or g\or h\or i\or j\or k\or l\or m\or n\or o\or p
\or q \or r \or s \or t \or u \or v \or w \or x \or y
\or z \else .\nlet \fi}
\def\asubsec#1{\smallskip{\bf\centerline{(\numblet) #1}\smallskip}}

\newcount\rslet\rslet=0
\def\romslet{\relax \global\advance\rslet by 1
\ifcase \rslet \ \or i\or ii\or iii\or iv\or v\or
vi\or vii\or viii\or ix\or x\or xi\or xii \else .\rslet \fi}

\newcount\rlet\rlet=0
\def\romlet{\relax \global\advance\rlet by 1
\ifcase \rlet \ \or I\or II\or III\or IV\or V\or
VI\or VII\or VIII\or IX\or X\or XI\or XII\or XIII\or XIV\or XV\or XVI
\or XVII \else .\rlet \fi}
\def\rchap#1{\bigskip\centerline{\bf{\romlet}. #1}\smallskip\noindent}

\def\pd#1#2{{\partial #1\over\partial #2}}      
\def\p2d#1#2{{\partial^2 #1\over\partial #2^2}} 
\def\td#1#2{{d #1\over d #2}}      
\def\t2d#1#2{{d^2 #1\over d #2^2}} 
\def\av#1{\langle #1\rangle}                    


\def\ith{{$i^{\rm th}$}}

\def\2kth{{$2k^{\rm th}$}}

\def\n-th{{$(n-1)^{\rm th}$}}

\def\N-th{{$(N-1)^{\rm th}$}}

\def\0th{$0^{\rm th}$}
\def\1st{$1^{\rm st}$}
\def\2nd{$2^{\rm nd}$}
\def\3rd{$3^{\rm rd}$}
\def\4th{$4^{\rm th}$}
\def\5th{$5^{\rm th}$}
\def\5th{$6^{\rm th}$}
\def\6th{$7^{\rm th}$}
\def\7th{$7^{\rm th}$}
\def\8th{$8^{\rm th}$}
\def\9th{$9^{\rm th}$}

\def\a{{\alpha}}
\def\b{{\beta}}
\def\g{\gamma}
\def\d{\delta}\def\D{\Delta}
\def\e{\epsilon}

\def\l{\lambda}
\def\m{\mu}
\def\n{\nu}

\def\t{\tau}

%% file: jour.tex
\def\aam #1 #2 #3 {{\sl Adv.\ Appl.\ Mech.} {\bf #1}, #2 (#3)}
\def\aap #1 #2 #3 {{\sl Adv.\ Appl.\ Prob.} {\bf #1}, #2 (#3)}
\def\ac #1 #2 #3 {{\sl Adv.\ Catal.} {\bf #1}, #2 (#3)}
\def\aces #1 #2 #3 {{\sl Adv.\ Chem.\ Eng.\ Sci.} {\bf #1}, #2 (#3)}
\def\acp #1 #2 #3 {{\sl Adv.\ Chem.\ Phys.} {\bf #1}, #2 (#3)}
\def\ae #1 #2 #3 {{\sl Ann.\ Eugenics} {\bf #1}, #2 (#3)}
\def\ah #1 #2 #3 {{\sl Adv.\ Hydrosci.} {\bf #1}, #2 (#3)}
\def\aichej #1 #2 #3 {{\sl AICHE J.} {\bf #1}, #2 (#3)}
\def\aj #1 #2 #3 {{\sl Astrophys.\ J.} {\bf #1}, #2 (#3)}
\def\ajp #1 #2 #3 {{\sl Am.\ J.\ Phys.} {\bf #1}, #2 (#3)}
\def\ams #1 #2 #3 {{\sl Ann.\ Math.\ Stat.} {\bf #1}, #2 (#3)}
\def\ap #1 #2 #3 {{\sl Adv.\ Phys.} {\bf #1}, #2 (#3)}
\def\apa #1 #2 #3 {{\sl Appl.\ Phys.\ A} {\bf #1}, #2 (#3)}
\def\apc #1 #2 #3 {{\sl Appl.\ Catal.} {\bf #1}, #2 (#3)}
\def\annp #1 #2 #3 {{\sl Ann.\ Phys.\ (N.Y.)} {\bf #1}, #2 (#3)}
\def\annpr #1 #2 #3 {{\sl Ann.\ Probab.} {\bf #1}, #2 (#3)}
\def\arfm #1 #2 #3 {{\sl Ann.\ Rev.\ Fluid Mech.} {\bf #1}, #2 (#3)}
\def\arpc #1 #2 #3 {{\sl Ann.\ Rev.\ Phys.\ Chem.} {\bf #1}, #2 (#3)}
\def\astech #1 #2 #3 {{\sl Aerosol Sci.\ Tech.} {\bf #1}, #2 (#3)}
\def\bcdg #1 #2 #3 {{\sl Ber.\ Chem.\ Dtsch.\ Ges.} {\bf #1}, #2 (#3)}
\def\bio #1 #2 #3 {{\sl Biometrika} {\bf #1}, #2 (#3)}
\def\bj #1 #2 #3 {{\sl Biophys.\ J.} {\bf #1}, #2 (#3)}
\def\ces #1 #2 #3 {{\sl Chem.\ Engr.\ Sci.} {\bf #1}, #2 (#3)}
\def\cf #1 #2 #3 {{\sl Combust.\ and Flame} {\bf #1}, #2 (#3)}
\def\cmp #1 #2 #3 {{\sl Commun.\ Math.\ Phys.} {\bf #1}, #2 (#3)}
\def\cp #1 #2 #3 {{\sl Chem.\ Phys.} {\bf #1}, #2 (#3)}
\def\contp #1 #2 #3 {{\sl Contemp.\ Phys.} {\bf #1}, #2 (#3)}
\def\cpam #1 #2 #3 {{\sl Commun.\ Pure Appl.\ Math.} {\bf #1}, #2 (#3)}
\def\crev #1 #2 #3 {{\sl Catal.\ Rev.} {\bf #1}, #2 (#3)}
\def\crII #1 #2 #3 {{\sl C. R. Acad.\ Sci.\ Ser.\ II} {\bf #1}, #2 (#3)}
\def\eul #1 #2 #3 {{\sl Europhys.\ Lett.} {\bf #1}, #2 (#3)}
\def\ic #1 #2 #3 {{\sl Icarus} {\bf #1}, #2 (#3)}
\def\iec #1 #2 #3 {{\sl Ind.\ Eng.\ Chem.} {\bf #1}, #2 (#3)}
\def\ijf #1 #2 #3 {{\sl Int.\ J.\ of Fracture} {\bf #1}, #2 (#3)}
\def\ijmpA #1 #2 #3 {{\sl Int.\ J.\ Modern Phys. A} {\bf #1}, #2 (#3)}
\def\ijmpB #1 #2 #3 {{\sl Int.\ J.\ Modern Phys. B} {\bf #1}, #2 (#3)}
\def\ijss #1 #2 #3 {{\sl Int.\ J.\ Solids Structures} {\bf #1}, #2 (#3)}
\def\jam #1 #2 #3 {{\sl J.\ Appl.\ Mech.} {\bf #1}, #2 (#3)}
\def\jap #1 #2 #3 {{\sl J.\ Appl.\ Phys.} {\bf #1}, #2 (#3)}
\def\japr #1 #2 #3 {{\sl J.\ Appl.\ Prob.} {\bf #1}, #2 (#3)}
\def\jaes #1 #2 #3 {{\sl J.\ Aerosol.\ Sci.} {\bf #1}, #2 (#3)}
\def\jas #1 #2 #3 {{\sl J.\ Atmos.\ Sci.} {\bf #1}, #2 (#3)}
\def\jasa #1 #2 #3 {{\sl J. Acous.\ Soc.\ Amer.} {\bf #1}, #2 (#3)}
\def\jc #1 #2 #3 {{\sl J.\ Catal.} {\bf #1}, #2 (#3)}
\def\jce #1 #2 #3 {{\sl J.\ Chem.\ Educ.} {\bf #1}, #2 (#3)}
\def\jcis #1 #2 #3 {{\sl J.\ Colloid Interface Sci.} {\bf #1}, #2 (#3)}
\def\jcg #1 #2 #3 {{\sl J.\ Crystal Growth} {\bf #1}, #2 (#3)}
\def\jcp #1 #2 #3 {{\sl J.\ Chem.\ Phys.} {\bf #1}, #2 (#3)}
\def\jcompp #1 #2 #3 {{\sl J.\ Comp.\ Phys.} {\bf #1}, #2 (#3)}
\def\jdep #1 #2 #3 {{\sl J.\ de Physique I} {\bf #1}, #2 (#3)}
\def\jdepl #1 #2 #3 {{\sl J. de Physique Lett.} {\bf #1}, #2 (#3)}
\def\jetp #1 #2 #3 {{\sl Sov.\ Phys.\ JETP} {\bf #1}, #2 (#3)}
\def\jetpl #1 #2 #3 {{\sl Sov. Phys.\ JETP Letters} {\bf #1}, #2 (#3)}
\def\jes #1 #2 #3 {{\sl J. Electrochem.\ Soc.} {\bf #1}, #2 (#3)}
\def\jfi #1 #2 #3 {{\sl J. Franklin Inst.} {\bf #1}, #2 (#3)}
\def\jfm #1 #2 #3 {{\sl J. Fluid Mech.} {\bf #1}, #2 (#3)}
\def\jgr #1 #2 #3 {{\sl J.\ Geophys.\ Res.} {\bf #1}, #2 (#3)}
\def\jif #1 #2 #3 {{\sl J. Inst.\ Fuel} {\bf #1}, #2 (#3)}
\def\jmo #1 #2 #3 {{\sl J. Mod. Opt.} {\bf #1}, #2 (#3)}
\def\jmp #1 #2 #3 {{\sl J. Math. Phys.} {\bf #1}, #2 (#3)}
\def\jms #1 #2 #3 {{\sl J. Memb.\ Sci.} {\bf #1}, #2 (#3)}
\def\josaA #1 #2 #3 {{\sl J. Opt.\ Soc.\ Am.\ A} {\bf #1}, #2 (#3)}
\def\josaB #1 #2 #3 {{\sl J. Opt.\ Soc.\ Am.\ B } {\bf #1}, #2 (#3)}
\def\jpa #1 #2 #3 {{\sl J. Phys.\ A} {\bf #1}, #2 (#3)}
\def\jpc #1 #2 #3 {{\sl J. Phys.\ C} {\bf #1}, #2 (#3)}
\def\jpd #1 #2 #3 {{\sl J. Phys.\ D} {\bf #1}, #2 (#3)}
\def\jpchem #1 #2 #3 {{\sl J. Phys.\ Chem.} {\bf #1}, #2 (#3)}
\def\jps #1 #2 #3 {{\sl J. Polymer.\ Sci.} {\bf #1}, #2 (#3)}
\def\jpsj #1 #2 #3 {{\sl J. Phys.\ Soc. Jpn.} {\bf #1}, #2 (#3)}
\def\jpso #1 #2 #3 {{\sl J. Power Sources} {\bf #1}, #2 (#3)}
\def\jsc #1 #2 #3 {{\sl J. Sci.\ Comp.} {\bf #1}, #2 (#3)}
\def\jsp #1 #2 #3 {{\sl J. Stat.\ Phys.} {\bf #1}, #2 (#3)}
\def\jtb #1 #2 #3 {{\sl J. Theor.\ Biol.} {\bf #1}, #2 (#3)}
\def\kc #1 #2 #3 {{\sl Kinet.\ Catal.\ (USSR)} {\bf #1}, #2 (#3)}
\def\macro #1 #2 #3 {{\sl Macromolecules} {\bf #1}, #2 (#3)}
\def\mclc #1 #2 #3 {{\sl Mol.\ Cryst.\ Liq.\ Cryst.} {\bf #1}, #2 (#3)}
\def\mubm #1 #2 #3 {{\sl Moscow Univ.\ Bull.\ Math.} {\bf #1}, #2 (#3)}
\def\nat #1 #2 #3 {{\sl Nature} {\bf #1}, #2 (#3)}
\def\npa #1 #2 #3 {{\sl Nucl.\ Phys.\ A} {\bf #1}, #2 (#3)}
\def\npb #1 #2 #3 {{\sl Nucl.\ Phys. B} {\bf #1}, #2 (#3)}
\def\PA #1 #2 #3 {{\sl PAGEOPH} {\bf #1}, #2 (#3)}
\def\pA #1 #2 #3 {{\sl Physica A} {\bf #1}, #2 (#3)}
\def\pB #1 #2 #3 {{\sl Physica B} {\bf #1}, #2 (#3)}
\def\pBC #1 #2 #3 {{\sl Physica B \& C} {\bf #1}, #2 (#3)}
\def\pD #1 #2 #3 {{\sl Physica D} {\bf #1}, #2 (#3)}
\def\pams #1 #2 #3 {{\sl Proc.\ Am.\ Math.\ Soc.} {\bf #1}, #2 (#3)}
\def\pcps #1 #2 #3 {{\sl Proc.\ Camb.\ Philos.\ Soc.} {\bf #1}, #2 (#3)}
\def\pfA #1 #2 #3 {{\sl Phys.\ Fluids A} {\bf #1}, #2 (#3)}
\def\pla #1 #2 #3 {{\sl Phys.\ Lett. A} {\bf #1}, #2 (#3)}
\def\plb #1 #2 #3 {{\sl Phys.\ Lett. B} {\bf #1}, #2 (#3)}
\def\pmB #1 #2 #3 {{\sl Philos.\ Mag. B} {\bf #1}, #2 (#3)}
\def\pnas #1 #2 #3 {{\sl Proc.\ Natl.\ Acad.\ Sci.} {\bf #1}, #2 (#3)}
\def\pr #1 #2 #3 {{\sl Phys.\ Rev.} {\bf #1}, #2 (#3)}
\def\pra #1 #2 #3 {{\sl Phys.\ Rev.\ A} {\bf #1}, #2 (#3)}
\def\prb #1 #2 #3 {{\sl Phys.\ Rev.\ B} {\bf #1}, #2 (#3)}
\def\prc #1 #2 #3 {{\sl Phys.\ Rev.\C} {\bf #1}, #2 (#3)}
\def\prd #1 #2 #3 {{\sl Phys.\ Rev.\ D} {\bf #1}, #2 (#3)}
\def\pre #1 #2 #3 {{\sl Phys.\ Rev.\ E} {\bf #1}, #2 (#3)}
\def\prept #1 #2 #3 {{\sl Phys.\ Repts.} {\bf #1}, #2 (#3)}
\def\prk #1 #2 #3 {{\sl Prog.\ Reac.\ Kinetics} {\bf #1}, #2 (#3)}
\def\prl #1 #2 #3 {{\sl Phys.\ Rev.\ Lett.} {\bf #1}, #2 (#3)}
\def\prsl #1 #2 #3 {{\sl Proc.\ Roy.\ Soc.\ London Ser. A} {\bf #1}, #2 (#3)}
\def\pss #1 #2 #3 {{\sl Prog.\ Surf.\ Sci.} {\bf #1}, #2 (#3)}
\def\pt #1 #2 #3 {{\sl Phys.\ Today} {\bf #1}, #2 (#3)}
\def\ptech #1 #2 #3 {{\sl Powder Tech.} {\bf #1}, #2 (#3)}
\def\ptrs #1 #2 #3 {{\sl Phil.\ Trans.\ Roy.\ Soc., Ser. A} {\bf #1}, #2 (#3)}
\def\rjpc #1 #2 #3 {{\sl Russ.\ J. Phys.\ Chem.} {\bf #1}, #2 (#3)}
\def\rmp #1 #2 #3 {{\sl Rev.\ Mod.\ Phys.} {\bf #1}, #2 (#3)}
\def\rpp #1 #2 #3 {{\sl Rep.\ Prog.\ Phys.} {\bf #1}, #2 (#3)}
\def\sci #1 #2 #3 {{\sl Science} {\bf #1}, #2 (#3)}
\def\sciam #1 #2 #3 {{\sl Scientific American} {\bf #1}, #2 (#3)}
\def\SIAMjam #1 #2 #3 {{\sl SIAM J.\ Appl.\ Math.} {\bf #1}, #2 (#3)}
\def\spu #1 #2 #3 {{\sl Sov.\ Phys.\ Usp.} {\bf #1}, #2 (#3)}
\def\SPEre #1 #2 #3 {{\sl SPE Reservoir Eng.} {\bf #1}, #2 (#3)}
\def\ssc #1 #2 #3 {{\sl Sol.\ State.\ Commun.} {\bf #1}, #2 (#3)}
\def\ss #1 #2 #3 {{\sl Surf.\ Sci.} {\bf #1}, #2 (#3)}
\def\stec #1 #2 #3 {{\sl Sov.\ Theor.\ Exp.\ Chem.} {\bf #1}, #2 (#3)}
\def\tAIME #1 #2 #3 {{\sl Trans.\ AIME} {\bf #1}, #2 (#3)}
\def\tfs #1 #2 #3 {{\sl Trans.\ Faraday Soc.} {\bf #1}, #2 (#3)}
\def\tpa #1 #2 #3 {{\sl Theor.\ Prob.\ Appl.} {\bf #1}, #2 (#3)}
\def\tpm #1 #2 #3 {{\sl Transport in Porous Media} {\bf #1}, #2 (#3)}
\def\usp #1 #2 #3 {{\sl Sov.\ Phys. -- Usp.} {\bf #1}, #2 (#3)}
\def\wrr #1 #2 #3 {{\sl Water Resources Res.} {\bf #1}, #2 (#3)}
\def\zpb #1 #2 #3 {{\sl Z. Phys.\ B} {\bf #1}, #2 (#3)}
\def\zpc #1 #2 #3 {{\sl Z. Phys.\ Chem.} {\bf #1}, #2 (#3)}
\def\ztf #1 #2 #3 {{\sl Zh. Tekh.\ Fiz.} {\bf #1}, #2 (#3)}
\def\zw #1 #2 #3 {{\sl Z. Wahrsch.\ verw.\ Gebiete} {\bf #1}, #2 (#3)}